\PassOptionsToPackage{hyphens}{url}

\documentclass[letterpaper,twocolumn,10pt]{article}
\usepackage{usenix-2020-09}

\usepackage{cite}
\usepackage{amsmath,amsfonts}
\usepackage{graphicx}
\usepackage{textcomp}
\usepackage{xcolor}
\usepackage{xspace}
\usepackage{booktabs}
\usepackage{enumitem}
\usepackage{framed}
\usepackage{changepage}
\usepackage{caption}
\usepackage{hyperref}
\usepackage{amssymb}
\usepackage{cleveref}
\usepackage{breakurl}           
\usepackage{url}                
\usepackage{tikz}
\usepackage{multirow}
\usepackage{soul}
\usepackage{url}

\usepackage{lipsum}

\usepackage{subcaption}
\usepackage{algorithm}
\usepackage[noend]{algpseudocode}

\usepackage{titling}

\def\BibTeX{{\rm B\kern-.05em{\sc i\kern-.025em b}\kern-.08em
T\kern-.1667em\lower.7ex\hbox{E}\kern-.125emX}}

\renewcommand{\paragraph}[1]{\vspace{.2\baselineskip}\noindent\textbf{#1.}}
\newcommand{\emphparagraph}[1]{\vspace{.1\baselineskip}\noindent\emph{\textbf{#1.}}}

\definecolor{formalshade}{rgb}{0.95,0.95,1}
\newenvironment{formal}{%
  \MakeFramed{\advance\hsize-\width\FrameRestore}%
  \noindent\hspace{-4.55pt}
  \begin{adjustwidth}{}{7pt}%
  \vspace{2pt}\vspace{2pt}%
}
{%
  \vspace{2pt}\end{adjustwidth}\endMakeFramed%
}

\definecolor{navy}{HTML}{2F729C}
\definecolor{graycanva}{HTML}{3B3B3B}

\hypersetup{
    colorlinks=true,
    linkcolor=graycanva,
    urlcolor=navy,
    citecolor=graycanva
}

\posttitle{\par\end{center}\vspace{-10pt}}

\newcommand{\addnoteauthor}[2]{%
  \expandafter\newcommand\csname #1\endcsname[1]{%
    {%
      \small\color{black}%
      \fbox{\scriptsize\bfseries\sffamily#1}%
      $\blacktriangleright${\sffamily\itshape\color{#2}##1}$\blacktriangleleft$%
      }%
    }%
  }

\crefformat{chapter}{\S#2#1#3}
\crefformat{section}{\S#2#1#3}
\crefformat{subsection}{\S#2#1#3}
\crefformat{subsubsection}{\S#2#1#3}

\newcommand{\smalltexttt}[1]{\texttt{\small{#1}}}

\usepackage{pifont}

\newcommand{\numlightcirclesans}[1]{\ding{\numexpr#1 + 191\relax}\xspace}
\newcommand{\numdarkcirclesans}[1]{\ding{\numexpr#1 + 201\relax}\xspace}

\makeatletter
\newcommand\footnoteref[1]{\protected@xdef\@thefnmark{\ref{#1}}\@footnotemark}
\makeatother

\newcommand\blfootnote[1]{%
  \begingroup
  \renewcommand\thefootnote{}\footnote{#1}%
  \addtocounter{footnote}{-1}%
  \endgroup
}

\definecolor{navy}{HTML}{2F729C}

\newcommand{\SYS}{\textsc{Padll}\xspace}
\newcommand{\Chef}{\textsc{Cheferd}\xspace}
\newcommand{\PAIO}{\textsc{Paio}\xspace}
\newcommand{\abci}{ABCI\xspace}
\newcommand{\abcifs}{\textsc{Pfs}$_{1}$\xspace}
\newcommand{\frontera}{Frontera\xspace}
\newcommand{\ALG}{PSFA\xspace}

\newcommand{\baseline}{\emph{\textbf{baseline}}\xspace}
\newcommand{\pass}{\emph{\textbf{passthrough}}\xspace}
\newcommand{\padll}{\emph{\textbf{padll}}\xspace}

\newcommand{\testbedA}{\emph{\textbf{A}}\xspace} 
\newcommand{\testbedB}{\emph{\textbf{B}}\xspace} 
\newcommand{\testbedC}{\emph{\textbf{C}}\xspace} 

\newcommand{\chef}{\textsc{Cheferd}\xspace}
\newcommand{\padllrepo}{\href{https://github.com/dsrhaslab/padll}{https://github.com/dsrhaslab/padll}}
\newcommand{\cheferdrepo}{\href{https://github.com/dsrhaslab/cheferd}{https://github.com/dsrhaslab/cheferd}}
\newcommand{\paiorepo}{\href{https://github.com/dsrhaslab/paio}{https://github.com/dsrhaslab/paio}}

\newcommand{\zenodo}{\href{https://zenodo.org/record/7627949/}{\textbf{https://zenodo.org/record/7627949/}}}

\usepackage{titling}
\addnoteauthor{rgmacedo}{blue}
\addnoteauthor{mmm}{green}
\addnoteauthor{jtpaulo}{red}
\addnoteauthor{yusuke}{gray}
\addnoteauthor{jason}{purple}
\addnoteauthor{amit}{orange}
\addnoteauthor{stephen}{black}
\addnoteauthor{todd}{yellow}

\begin{document}

\date{}
\title{\Large \bf \SYS: Taming Metadata-intensive HPC Jobs Through Dynamic,\\ Application-agnostic QoS Control} 



\author{
{\rm Ricardo\ Macedo$^{\varnothing\star}$}\\
{\normalsize INESC TEC \& U. Minho}
\and
{\rm Mariana\ Miranda$^{\varnothing}$}\\
{\normalsize INESC TEC \& U. Minho}
\and
{\rm Yusuke\ Tanimura}\\
{\normalsize AIST}
\and
{\rm Jason\ Haga}\\
{\normalsize AIST}
\and
{\rm Amit\ Ruhela}\\
{\normalsize TACC \& UT Austin}
\and
{\rm Stephen\ Lien\ Harrell}\\
{\normalsize TACC \& UT Austin}
\and
{\rm Richard\ Todd\ Evans}\\
{\normalsize Intel}
\and
{\rm José\ Pereira}\\
{\normalsize INESC TEC \& U. Minho}
\and
{\rm João\ Paulo}\\
{\normalsize INESC TEC \& U. Minho}
\vspace*{-25pt}
} 

\maketitle

\begin{abstract}
    Modern I/O applications that run on HPC infrastructures are increasingly becoming read and metadata intensive. 
    However, having multiple applications submitting large amounts of metadata operations can easily saturate the shared parallel file system's metadata resources, leading to overall performance degradation and I/O unfairness. 
    %
    We present \SYS, an application and file system agnostic storage middleware that enables QoS control of data and metadata workflows in HPC storage systems.
    It adopts ideas from Software-Defined Storage, building data plane stages that mediate and rate limit POSIX requests submitted to the shared file system, and a control plane that holistically coordinates how all I/O workflows are handled.
    We demonstrate its performance and feasibility under multiple QoS policies using synthetic benchmarks, real-world applications, and traces collected from a production file system.
    Results show that \SYS can enforce complex storage QoS policies over concurrent metadata-aggressive jobs, ensuring fairness and prioritization.
\end{abstract}
\section{Introduction}
\label{sec:introduction}

Modern supercomputers\blfootnote{$^{\varnothing}$Both authors contributed equally to this work.}\blfootnote{$^{\star}$Correponding author: \href{mailto:ricardo.g.macedo@inesctec.pt}{ricardo.g.macedo@inesctec.pt}} are establishing a new era in high-performance computing (HPC), providing unprecedented compute power that enables parallel applications to run at large-scale~\cite{frontier, aurora, Fugaku:2020:Sato}.
However, contrary to long-lived assumptions about HPC workloads where applications were predominately compute-bound and write-dominated, modern applications (\emph{e.g.,} Deep Learning (DL) training) are data-intensive, read-dominated, and generate massive bursts of metadata operations~\cite{DLIO:2021:Devarajan, CharacterizingDeepLearning:2018:Chien,RevisitingIOBehavior:2019:Patel}.
In fact, recent studies have noted that many applications spend 15-40\% of their execution time performing storage I/O, and expect this value to increase for exascale systems~\cite{GIFT:2020:Patel,CALCioM:2014:Dorier,PerformanceCharacterization:2017:Daley,RevisitingIOBehavior:2019:Patel}.

Contrary to compute resources (\emph{e.g.,} CPU, GPU), which are exclusively reserved to a given job, storage resources -- including both \emph{data} and \emph{metadata} -- are often shared across jobs, for example, when accessing the same Parallel File System (PFS). 
While modern workloads demand scalable, high throughput, and low latency storage, having multiple concurrent jobs competing for shared storage resources can lead to severe I/O contention and overall performance degradation~\cite{GIFT:2020:Patel,LustreTacc:2021:Liu,TBF:2017:Qian}.
Thus, not ensuring storage quality of service (QoS) guarantees in large-scale HPC systems means that jobs will unfairly access shared resources and execute without sustained I/O performance~\cite{TBF:2017:Qian}.


\paragraph{Challenges}
Efficiently controlling I/O workflows of large-scale HPC storage systems poses unique challenges, of which existing approaches have been unable to address.

\emphparagraph{Manual intervention}
In several HPC facilities, system administrators stop jobs with aggressive I/O behavior (\emph{e.g.,} accessing large datasets made of small-sized files, overloading the shared storage with unnecessary I/O requests) and temporarily suspend job submission access for users that do not comply with the cluster's guidelines~\cite{ContainersSupercomputer:2021:Huang, LustreTacc:2021:Liu}.
However, this \emph{reactive approach} is triggered when the job has already slowed the storage system and impacted the QoS of other jobs.

\emphparagraph{Intrusiveness to I/O layers}
While solutions like GIFT~\cite{GIFT:2020:Patel}, CALCioM~\cite{CALCioM:2014:Dorier}, and TBF~\cite{TBF:2017:Qian} aim at mitigating I/O contention and variability, these are tightly coupled to the implementation of core layers of the HPC I/O stack, including the shared file system, job scheduler, and I/O libraries.
Such an approach requires profound code refactoring, increasing the work needed to maintain and port it to new platforms.
For instance, optimizations made at Lustre may not be directly applicable over other file systems (\emph{e.g.,} BeeGFS, PVFS), as even though they share a similar high-level design, the internal I/O logic differs across implementations.

\emphparagraph{Partial visibility and I/O control}
Few solutions enable QoS control from the application-side (\emph{i.e.,} at the compute node level), thus not requiring changes to core layers of the I/O stack~\cite{OOOPS:2020:Huang}.
However, these act in isolation (\emph{i.e.,} agnostic of other jobs in execution), being unable to holistically coordinate the I/O generated from multiple jobs that compete for shared storage, ultimately leading to I/O contention and waste of system resources (\emph{e.g.,} unused I/O bandwidth)~\cite{PAIO:2022:Macedo,Pisces:2012:Shue}.

\emphparagraph{Metadata remains overlooked}
While existing proposals focus on achieving QoS over data workflows (I/O bandwidth)~\cite{GIFT:2020:Patel,CALCioM:2014:Dorier,IOrchestrator:2010:Zhang,UShape:2011:Zhang,SchedulingUnderCongestion:2015:Gainaru,TBF:2017:Qian,SIREN:2018:Karki,MappingScheduling:2020:Carretero}, the metadata counterpart has not received the same level of attention.
In fact, several HPC centers are observing a surge of metadata operations in their clusters and expect this to become more severe over time.
This is problematic given that even the I/O operations of a single job can saturate the PFS metadata resources, leading to unresponsiveness of the file system and increased execution time for all running jobs~\cite{ContainersSupercomputer:2021:Huang,LustreTacc:2021:Liu}.

\paragraph{This work}
To address these challenges, we present \SYS, an application and file system agnostic storage middleware that enables QoS control of data and metadata workflows in HPC storage systems.
Fundamentally, it allows system administrators to proactively and holistically control the rate at which POSIX requests are submitted to the PFS.

\SYS adopts ideas from Software-Defined Storage~\cite{SDSsurvey:2020:Macedo}, following a decoupled design that separates the I/O logic into a \emph{data plane} and a \emph{control plane}.
The \emph{data plane} is a multi-stage component that actuates at the compute node level, where each stage mediates the I/O requests between a given application and the shared file system. 
Specifically, stages transparently handle applications' requests by intercepting POSIX calls (\emph{e.g.,} \texttt{open}, \texttt{close}, \texttt{read}, \texttt{getattr}) and dynamically rate limiting those that are destined towards the PFS. 
This makes \SYS applicable over multiple applications and cross-compatible with POSIX-compliant file systems, without requiring changes to any core layer of the HPC I/O stack.

Stages are then controlled by a logically centralized manager, the \emph{control plane}, that defines how all I/O workflows should be handled.
It acts as a global coordinator with system-wide visibility that continuously monitors and adjusts the I/O rate of data plane stages.
It does so by dynamically allocating storage resources (\emph{i.e.,} metadata rate, I/O bandwidth) among jobs upon workload and system variations, ensuring that QoS policies are met at all times.
To orchestrate a large number of data plane stages concurrently, the control plane is hierarchically distributed, made of \emph{global} and \emph{local controllers}.

To ensure custom and fine-grained control over I/O workflows, \SYS enables system administrators to specify QoS policies through \emph{control algorithms}, which can be as simple as statically rate limiting a specific type of request (\emph{e.g.,} \texttt{open}) of a given job, to more complex ones, as achieving proportional sharing of metadata resources across all active jobs~\cite{NetworkCalculus:2001:Boudec,DRF:2011:Ghodsi}.

\paragraph{Implementation and evaluation}
To validate the performance and feasibility of our approach, we implemented a \SYS prototype, including multiple control algorithms to enforce different storage QoS policies, namely \emph{uniform} and \emph{priority-based rate distributions}, \emph{proportional sharing}, and a new \emph{max-min fair share} algorithm that prevents over-provisioning under volatile workloads.

Experiments were conducted using both synthetic benchmarks and real-world applications (IOR~\cite{IOR:2007:Shan} and TensorFlow~\cite{TensorFlow:2016:Abadi}, respectively), as well as traces of metadata operations collected from a production Lustre file system of the \abci supercomputer.
Results demonstrate that:
(1) \SYS effectively controls the rate of I/O workflows at different granularities, including request type (\emph{e.g.,} \texttt{open}, \texttt{read}, \texttt{getattr}), request class (\emph{e.g.,} metadata, data), and job;
(2) it enables enforcing storage QoS policies over distributed, metadata-aggressive jobs holistically; 
(3) when configured with our new control algorithm, under volatile workloads, \SYS maximizes the use of metadata resources, accelerating the performance of resource-hungry jobs without degrading over-provisioned ones;
and (4) a single stage is able to service requests at high throughput rates (up to 3.20~Mops/s), and the control plane can manage the overall system at $\mu$$s$-scale.

\vspace{.2\baselineskip}\noindent
In summary, the paper makes the following contributions:

\begin{itemize}[nosep]
    \item A \textbf{study} that \emph{analyzes traces} from a production Lustre file system at \abci, highlighting the importance of ensuring QoS over metadata resources (\cref{sec:background}).
    \item \textbf{\SYS}, an application and PFS agnostic storage middleware that enables QoS control in HPC storage systems (\cref{sec:padll}).
    The system is publicly available at \href{https://github.com/dsrhaslab/padll}{dsrhaslab/padll}, \href{https://github.com/dsrhaslab/cheferd}{dsrhaslab/cheferd}, and Zenodo~\cite{Padll:2023:Macedo} repositories.
    \item A \textbf{new max-min fair share algorithm} that enables differentiated QoS across multiple jobs, while preventing resource over-provisioning under volatile workloads (\cref{sec:control-algorihtms}).
    \item \textbf{Experimental results} demonstrating \SYS's per\-for\-man\-ce and applicability under different scenarios using both synthetic and realistic I/O workloads (\cref{sec:evaluation}).
\end{itemize}

\section{Background and Motivation}
\label{sec:background}

Parallel file systems are the storage backbone of HPC infrastructures, being used to store and retrieve, on a daily basis, petabytes of data from hundreds to thousands of concurrent jobs.
In this paper, we focus on Lustre-like file systems (\emph{e.g.,} Lustre~\cite{Lustre:2003:Schwan}, BeeGFS~\cite{BeeGFS:2019:Chowdhury}, PVFS~\cite{PVFS:2000:Carns}), which are present in most TOP500 supercomputers.
A typical Lustre-like file system consists of several building blocks.
\emph{Me\-ta\-da\-ta Servers (MDSs)} maintain the file system namespace (\emph{e.g.,} file names and layouts, permissions, extended attributes) and handle all metadata operations.
The namespace is persisted in a single or multiple \emph{Metadata Targets (MDTs)} nodes.
Data operations are serviced by \emph{Object Storage Servers (OSSs)} which are connected to compute nodes via high-speed interconnects, and store files on \emph{Object Storage Targets (OSTs)}.
Files are typically distributed across multiple OSTs for parallelism and availability.
File system \emph{clients} reside at compute nodes (in kernel-level) and access the file system using standard POSIX system calls (\emph{e.g.,} \texttt{open}, \texttt{read}, \texttt{close}, \texttt{getattr}).

Depending on the scale of the file system, metadata nodes assume different configurations~\cite{LustreMDS}.
In some deployments, the namespace is persisted across multiple MDTs and a single MDS handles all metadata operations, having additional MDS nodes as standby replicas; in others, different MDSs/MDTs manage/persist different parts of the namespace.

\paragraph{Metadata workflow and limitations}
Regardless of the application, workload, or job, whenever a file needs to be accessed (\emph{e.g.,} create/open/remove file, access control, extended attributes) the main I/O path always flows through the metadata service.
When creating files, the file system client issues a RPC routine to the MDS, which will create a new entry in the namespace and assign OSTs in a capacity-balanced manner to persist the data; for existing files, the MDS retrieves information about the file stripe and OST mappings.

When used at scale, this centralized design comprises several limitations that can severely bottleneck the file system and impact the performance of all running jobs.
First, different metadata operations carry different costs to the PFS.
Depending on the file system implementation, read-only operations such as \texttt{getattr} only require acquiring read-locks, while operations like \texttt{open}, \texttt{close}, and \texttt{unlink} require more expensive locking, as they need to update the namespace state~\cite{LustreArchitecture:2019:Braam,LustreMDC}.
Other operations, such as \texttt{mkdir} or \texttt{rename}, require even stronger guarantees (\emph{i.e.,} \emph{atomicity}).
Second, modern workloads, such as DL training, comprise large-scale datasets that can reach TiB in size and are made of multiple small-sized files, which generate high and continuous bursts of metadata operations~\cite{FMA:2016:Defferrard,OpenImages:2020:Kuznetsova}.
Third, the number of file system \emph{clients} is several times higher than available MDSs, which can become saturated when several concurrent jobs have aggressive I/O metadata behavior~\cite{LustreTacc:2021:Liu}. 


\begin{figure}[t]
    \centering
    \includegraphics[width=1\linewidth]{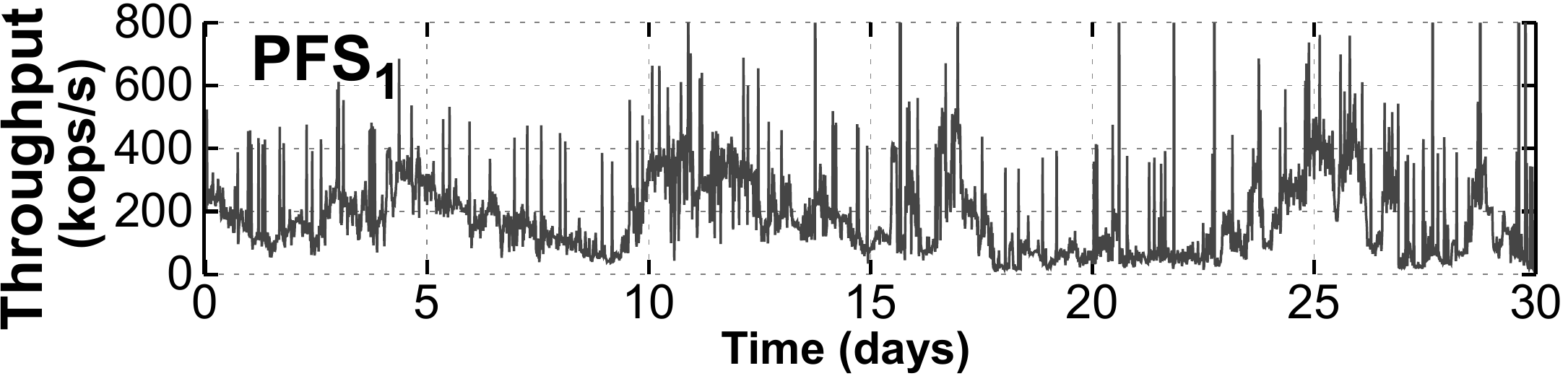}
    \vspace*{-15pt}
    \caption{Metadata throughput in \abcifs over a 30-day period.}
    \label{fig:abci-overall-metadata-ops}
\end{figure}

\subsection{Analyzing Metadata Operations in Production Clusters}
\label{subsec:study}

To understand the impact of metadata operations in production, we analyze the logs of a Lustre file system from the \abci supercomputer~\cite{abci}.
The file system is a DDN ExaScaler Lustre composed of 2~MDSs in a hot-standby configuration, backed by 6~MDTs, and 36~OSTs that provide 9.5~PiB of storage capacity. 
We refer to this file system as \abcifs.

We monitored the I/O activity of the most frequent metadata operations at MDSs/MDTs, using DDNStorage's LustrePerfMon~\cite{LustrePerfMon}.
We collected per-MDT performance statistics for \texttt{open}, \texttt{clo\-se}, \texttt{get\-at\-tr}, \texttt{set\-at\-tr}, \texttt{re\-na\-me}, \texttt{mk\-dir}, \texttt{mk\-nod}, \texttt{rm\-dir}, \texttt{stat\-fs}, \texttt{sync}, and \texttt{un\-li\-nk} operations. 
The logs report per-operation performance statistics captured with 1-minute samples over a 30-day observation period.
Further, we also monitored the I/O bandwidth (\texttt{read} and \texttt{write}) observed by OSSs over the same collection period.


\paragraph{Overall metadata load}
We first examine the throughput of metadata operations throughout the overall observation period.
Fig.~\ref{fig:abci-overall-metadata-ops} depicts the rate of all collected metadata operations at \abcifs.
Metadata operations are submitted at a massive rate, averaging 200~kops/s.
Over different periods, \abcifs continuously serves requests over 400~kops/s, which last several hours to days, and experiences bursts that peak at 1~Mops/s.
Indeed, the workload is extremely volatile, frequently exhibiting periods of low throughput (50~kops/s or lower) to immediately spike up to 450~kops/s (or higher).

Furthermore, we observe that this load is much higher than those reported in other clusters~\cite{RevisitingIOBehavior:2019:Patel}. 
For example, a study from NERSC reports that the PFS shared by \emph{Edison} and \emph{Cori} supercomputers had an average rate of 9.7~kops/s and 7~kops/s for \texttt{open} and \texttt{close} operations, respectively; while \abcifs experiences 29~kops/s and 43.5~kops/s.
While the metadata load depends on different factors (\emph{e.g.,} cluster size, workload, file organization), we suspect that these values mainly stem from the type of jobs conducted at \abci, which are mostly AI-oriented (\emph{e.g.,} DL training).

\begin{formal}
    \emph{\textbf{Observation:} Modern I/O workloads generate massive amounts of metadata operations.
    Based on previous studies~\cite{RevisitingIOBehavior:2019:Patel} and the results observed from \abcifs, it is expected that these values will continue to increase over time.
    This means that exclusively ensuring QoS over data workflows (i.e., \texttt{read} and \texttt{write}) is no longer enough to achieve fair access to PFS resources, and thus, metadata operations need to be managed as well.} 
\end{formal}
    

\begin{figure}[t]
    \centering
    \includegraphics[width=1\linewidth]{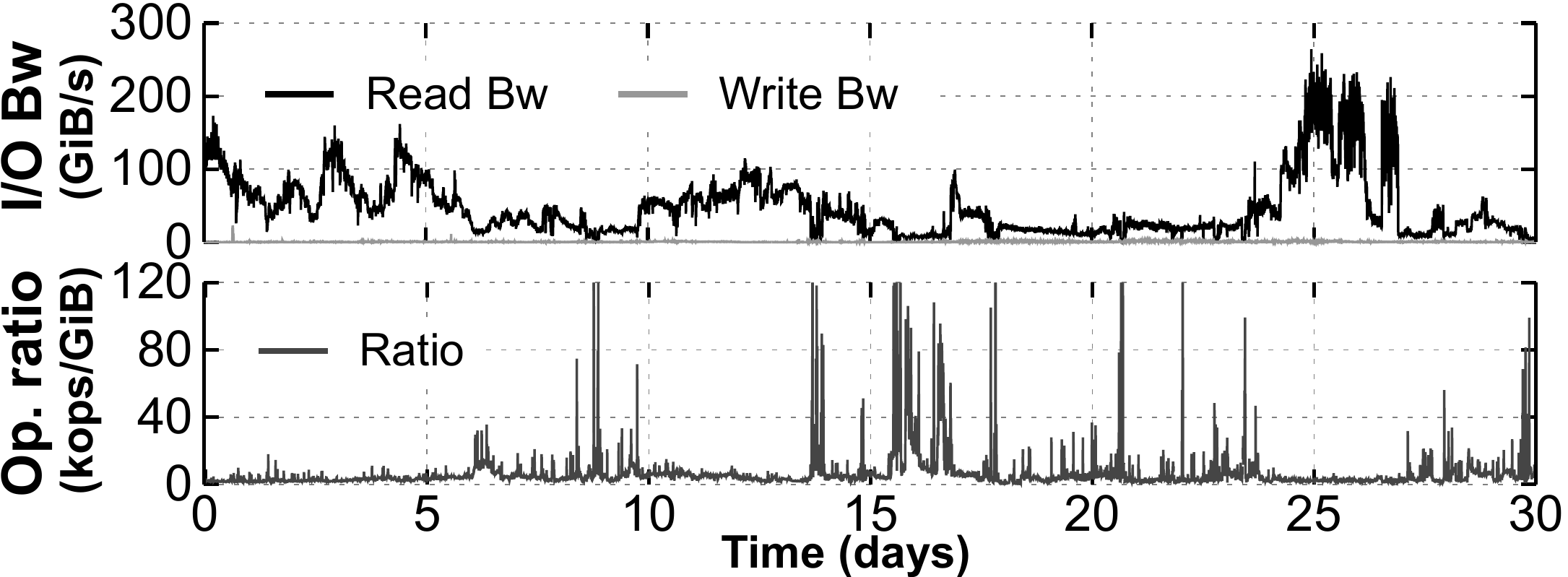}
    \vspace*{-15pt}
    \caption{Ratio of metadata operations to I/O bandwidth in \abcifs.}
    \label{fig:abci-oss-bandwidth}
\end{figure}

\paragraph{Ratio of metadata operations to I/O bandwidth}
We now analyze the correlation between data and metadata operations at \abcifs.
As most jobs conducted at \abci are AI-oriented, \abcifs experiences low write throughput (average rate of 0.6~GiB/s), while reads are served at an average rate of 48~GiB/s, as depicted in Fig.~\ref{fig:abci-oss-bandwidth} (top).
However, comparing the amount of metadata operations and the I/O bandwidth serviced by \abcifs, as depicted in Fig.~\ref{fig:abci-oss-bandwidth} (bottom), we observe that, in several periods, metadata operations have significantly higher throughput than GiB/s read/written from/to the PFS.
For instance, between days 13 and 20, metadata operations were submitted at a rate over 120~kops for each GiB (or 120 ops/MiB) read/written from/to the PFS.
This means that even if hard QoS limits are imposed over data operations, metadata workflows may still remain unchanged.

\begin{formal}
    \emph{\textbf{Observation:} We observed several periods where the amount of submitted metadata operations far exceeds the GiBs of data read/written from/to \abcifs.
    This means that there is not a strict dependency between both operation classes, 
    consolidating the need for ensuring QoS over metadata workflows as well.}
\end{formal}


\begin{figure}[t]
    \centering
    \includegraphics[width=1\linewidth]{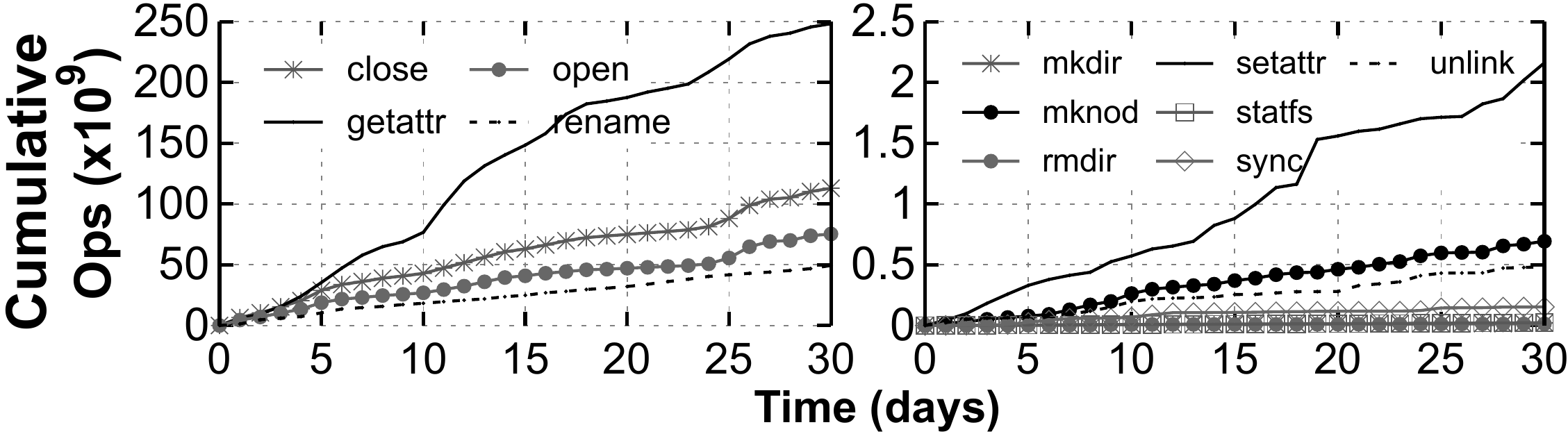}
    \vspace*{-15pt}
    \caption{Cumulative metadata operations in \abcifs.}
    \label{fig:abci-cumulative-metadata}
\end{figure}

\paragraph{Type and frequency of metadata operations}
Fig.~\ref{fig:abci-cumulative-metadata} shows the type and amount of metadata operations in \abcifs.
\texttt{open}, \texttt{close}, \texttt{getattr}, and \texttt{rename} are the most frequent operations, accounting for $98\%$ of the total load.
Notoriously, several of these are particularly costly to the PFS and prone to cause I/O contention, due to expensive locking (\emph{i.e.,} \texttt{open} and \texttt{close}) and atomicity guarantees (\emph{i.e.,} \texttt{rename}).
As for \texttt{getattr} operations, while less costly, \abcifs received almost 250 billion requests throughout the overall observation period (corresponding to $\approx$47\% of the total load), representing an average and continuous rate of 95.8~kops/s.

\begin{formal}
    \emph{\textbf{Observation:} The most predominant metadata operations, na\-me\-ly \texttt{open}, \texttt{close}, \texttt{rename}, and \texttt{getattr}, account for 98\% of \abcifs metadata load.
    Given that not all operations entail the same cost and I/O pressure over the shared metadata resources, operations should be controlled with fine-granularity.
    }
\end{formal}

\section{\SYS Storage Middleware}
\label{sec:padll}

\SYS is a storage middleware that enables system administrators to proactively and holistically control the rate of data and metadata workflows to achieve QoS in HPC storage systems. 
Its design is built under the following core principles.

\paragraph{Application and PFS agnostic}
\SYS does not require code changes to any core layer of the HPC I/O stack, being ag\-nos\-tic of the applications it is controlling as well the file system to which requests are submitted to.
This makes \SYS applicable over multiple applications and compatible with PO\-SIX-compliant storage systems, including both local (\emph{e.g.,} \smalltexttt{xfs}, \smalltexttt{ext4}) and distributed file systems (\emph{e.g.,} Lustre, BeeGFS).

\paragraph{Fine-grained I/O control}
\SYS classifies, differentiates, and controls requests at different levels of granularity, including \emph{operation type} (\emph{e.g.,} \smalltexttt{open}, \smalltexttt{read}), \emph{operation class} (\emph{e.g.,} \smalltexttt{data}, \smalltexttt{metadata}), \emph{user}, and \emph{job}, which allows applying different types of policies (\emph{e.g.,} only rate limit \smalltexttt{open} calls, rate limit all \smalltexttt{metadata} operations, rate limit job \texttt{xyz} to $X$ ops/s).

\paragraph{Global visibility}
\SYS ensures holistic control of all I/O workflows and coordinated access to the PFS, preventing I/O contention and unfair usage of shared storage resources.

\paragraph{Custom QoS specification}
\SYS enables system administrators to create custom QoS policies for rate limiting jobs running at the cluster (\emph{e.g.,} uniform~\cite{OOOPS:2020:Huang} and priority-based rate distribution~\cite{TBF:2017:Qian}, proportional sharing~\cite{IOFlow:2013:Thereska,PAIO:2022:Macedo}, DRF~\cite{DRF:2011:Ghodsi}), protecting the PFS from greedy jobs and I/O burstiness.

\begin{figure}[t]
    \centering
    \includegraphics[width=1\linewidth]{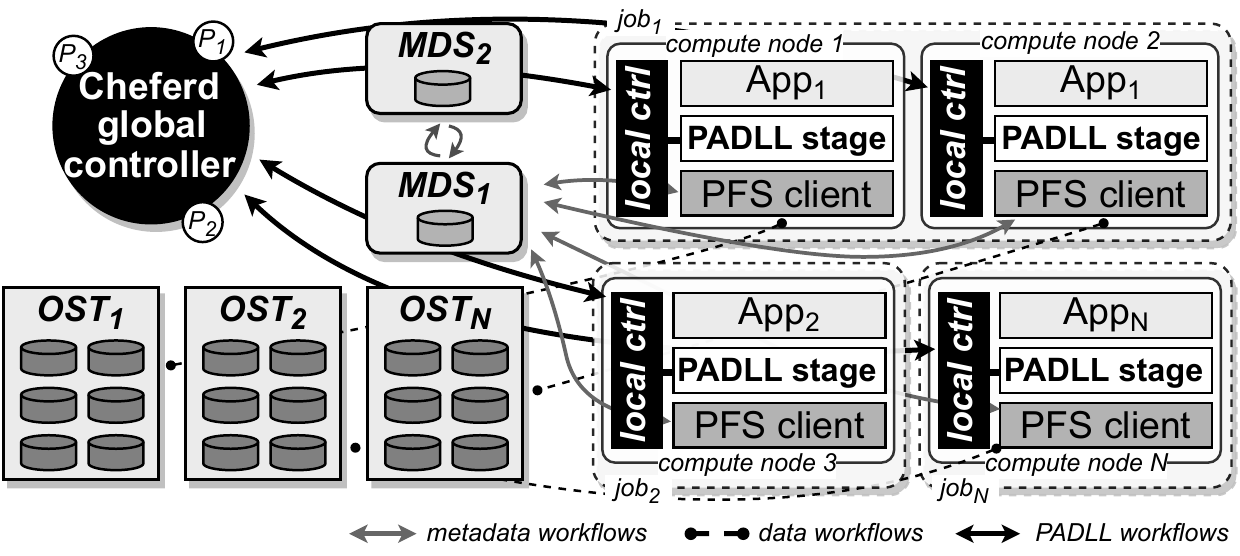}
    \vspace*{-15pt}
    \caption{\textbf{\SYS high-level architecture.} It is composed of a hierarchical control plane (\Chef) and multiple data plane stages.}
    \label{fig:padll-high-level-arch}
\end{figure}

\vspace*{5pt}
Fig.~\ref{fig:padll-high-level-arch} outlines \SYS's high-level architecture.
It follows a decoupled design that separates the I/O logic into two planes of functionality. 
The \emph{data plane} (\cref{subsec:data-plane}) is a multi-stage component that provides the building blocks for differentiating and rate limiting I/O workflows.
The \emph{control plane} (\cref{subsec:control-plane}) is a global coordinator that manages all data plane stages to ensure that storage QoS policies are met over time and adjusted according to workload variations.


\subsection{Data Plane}
\label{subsec:data-plane}

\SYS's data plane stages (or \emph{stages} for short) actuate at the compute node level, each sitting between the application and the PFS.
\SYS transparently intercepts (using \smalltexttt{LD\_PRELOAD}) and reimplements multiple POSIX system calls from different operation classes before being submitted to the PFS, including \emph{data} (\emph{e.g.,} \texttt{read}, \texttt{pwrite}), \emph{metadata} (\emph{e.g.,} \texttt{open}, \texttt{rename}), \emph{extended attributes} (\emph{e.g.,} \texttt{getattr}, \texttt{setattr}), and \emph{directory management} (\emph{e.g.,} \texttt{mkdir}, \texttt{mknod}).

To control the rate of I/O workflows of a given job, multiple \SYS stages may be used. 
Under single node jobs, a single data plane stage controls all I/O workflows. As depicted in Fig.~\ref{fig:padll-high-level-arch}, this is the case of $job_2$ where App$_2$ only executes at \emph{compute node 3}.
For distributed jobs, where application instances run on separate compute nodes, 
multiple data plane stages need to be set (\emph{i.e.,} one per instance). 
For example, as depicted for $job_1$ in Fig.~\ref{fig:padll-high-level-arch}, two stages are required to effectively rate limit all I/O workflows of App$_1$, since it executes in \emph{compute nodes 1} and \emph{2}.
Furthermore, each compute node can also have multiple stages, created from multi-process applications (\emph{i.e.,} one stage per process).

\begin{figure}[t]
    \centering
    \includegraphics[width=1\linewidth]{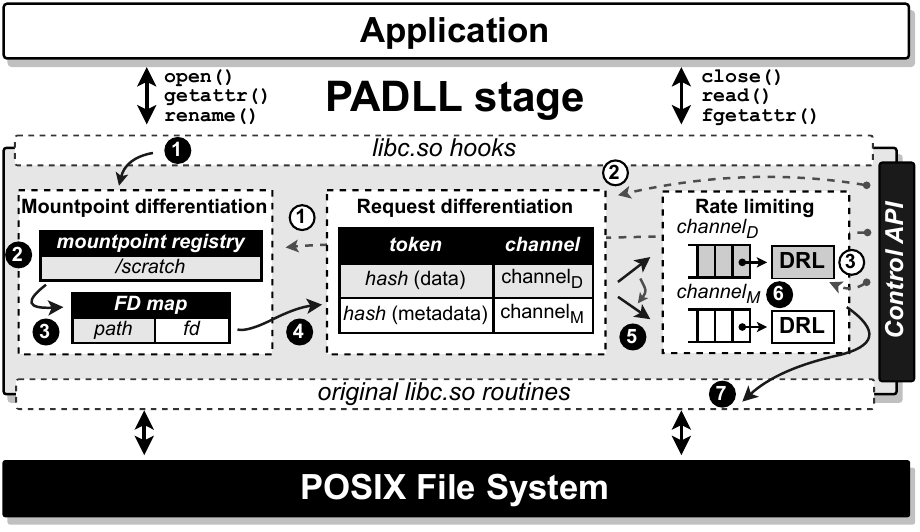}
    \vspace*{-17.5pt}
    \caption{\textbf{\SYS data plane stage design and operation flow.}
    Black circles depict the operation flow of intercepted POSIX requests, while white circles depict housekeeping and control operations of \SYS.}
    \label{fig:padll-stage-design}
\end{figure}

To handle the I/O workflows of a given application, stages are organized in three main components, as depicted in Fig.~\ref{fig:padll-stage-design}.

\paragraph{Mountpoint differentiation}
Internally, compute nodes hold multiple file systems, including \emph{local}, which are used for managing local storage devices (\emph{e.g.,} \smalltexttt{xfs}, \smalltexttt{tmpfs}), and \emph{remote}, for accessing files in distributed storage systems like Lustre and BeeGFS. 
Given that \SYS intercepts POSIX requests regardless of the targeted file system, it needs to identify which requests are destined towards the PFS, so these can be treated accordingly.
This is achieved in three phases.

\vspace*{2.5pt}\noindent\emph{Registering mountpoints:}
first, the system administrator defines which mountpoints should be managed with \SYS, by registering their full path on a mountpoint registry (\numlightcirclesans{1}). 
For example, as depicted in Fig.~\ref{fig:padll-stage-design}, the stage only handles the requests that are destined towards \smalltexttt{/scratch}.

\vspace*{2.5pt}\noindent\emph{Handling path-based operations:}
all system calls that define the \emph{pathname} of the targeted file, such as \texttt{open}, \texttt{fopen}, \texttt{rename}, and \texttt{mkdir}, are then intercepted (\numdarkcirclesans{1}) and analyzed (\numdarkcirclesans{2}). 
Requests that are destined towards the registered mountpoints proceed to the subsequent components (\numdarkcirclesans{4}); otherwise, these are directly submitted to the corresponding file system without additional processing (\numdarkcirclesans{7}).

\vspace*{2.5pt}\noindent\emph{Handling file descriptor (FD) based operations:}
to determine if a system call that accesses files through FDs (\emph{e.g.,} \texttt{read}, \texttt{fgetattr}) is destined towards a registered mountpoint, for each \texttt{open}-based call, \SYS stores the resulting FD in a file mapping module (\numdarkcirclesans{3}). 
Whenever any of these system calls is intercepted (\numdarkcirclesans{1}), \SYS verifies if the corresponding FD is valid (\numdarkcirclesans{3}), proceeding to the subsequent components (\numdarkcirclesans{4}); otherwise, it is submitted to the corresponding file system without changes (\numdarkcirclesans{7}).
On \texttt{close}, the FD is removed from the file mapping.
File pointer based operations (\emph{e.g.,} \texttt{fread}) are handled in a similar manner.

\paragraph{Request differentiation and rate limiting}
Internally, as depicted in Fig.~\ref{fig:padll-stage-design}, stages are organized in multiple queues (entitled as \emph{channels}), each with a token-bucket that determines the rate of its requests (entitled as \emph{DRL}).
A token-bucket is a commonly used mechanism for controlling the rate and burstiness of I/O workflows~\cite{NetworkCalculus:2001:Boudec}.
Each of these \emph{channels} only serves a specific set of requests.
For example, \emph{channel}$_D$ and \emph{channel}$_M$ handle all \texttt{data} and \texttt{metadata} operations destined towards \smalltexttt{/scratch}, respectively.
The type of requests each \emph{channel} handles (\numlightcirclesans{2}), as well as the rate of each token-bucket are set by the control plane (\numlightcirclesans{3}).

After validating their associated mountpoint, requests are differentiated based on a specific set of attributes that characterize them, including the \emph{operation type} (\emph{e.g.,} \texttt{open}, \texttt{get\-attr}), \emph{operation class} (\emph{e.g.,} \texttt{metadata}, \texttt{data}), \emph{operation size}, \emph{userID}, and \emph{jobID} (\numdarkcirclesans{4}).
For each request, the stage hashes its attributes into a fixed-size token through a computationally cheap hashing scheme~\cite{MurmurHash:2020:Appleby}, which maps the request to the cor\-res\-pon\-ding \emph{channel} that will enforce it (\numdarkcirclesans{5}).
If no match is found, it means that the request should not be handled by \SYS, being submitted to the file system without additional processing (\numdarkcirclesans{7}). 
Once in the \emph{channel}, requests are then processed (\emph{i.e.,} dequeued) and rate limited according to the token-bucket's rate (\numdarkcirclesans{6}).
After this process, requests are then submitted to the targeted POSIX file system (\numdarkcirclesans{7}).


\subsection{Hierarchical Control Plane}
\label{subsec:control-plane}

The control plane is a logically centralized component with system-wide visibility that defines how all I/O workflows in the HPC cluster are handled.
It does so by continuously communicating with data plane stages to \emph{collect I/O metrics} (\emph{e.g.,} operation rate, I/O bandwidth) and \emph{enforce stage-specific rules} that dynamically adjust the workflow's rate (\emph{i.e.,} at token-buckets) to respond to workload and system variations, as well as new policies set by system administrators.
 
As depicted in Fig.~\ref{fig:padll-high-level-arch}, \SYS's control plane follows a hierarchical distribution.
\emph{Local controllers} have local visibility and manage all data plane stages of a given compute node.
A \emph{global controller} has global visibility and enforces cluster-wide QoS policies by orchestrating all \emph{local controllers}.
Since multiple stages can execute in the same compute node, this hierarchy of controllers allows minimizing the number of connections and exchanged messages to the \emph{global controller}. 

\paragraph{Control logic}
In \SYS, the control logic is specified through storage QoS policies. 
These can be as simple as individually set the rate for the \texttt{open} calls of a given job, to more complex ones such as dynamically reserve shares of metadata operations for all jobs in the cluster.
The latter are defined through \emph{control algorithms}, which are implemented in a \emph{feedback control loop}, where the control plane repeatedly performs four main steps, namely \emph{collect}, \emph{compute}, \emph{enforce}, and \emph{sleep}.

\vspace*{.1\baselineskip}\noindent\emph{Collect:}
\emph{local controllers} continuously collect statistics (\emph{e.g.,} per-stage metadata rate) from their assigned data plane stages (\emph{i.e.,} which are co-located in the same compute node). 
These metrics are aggregated and reported to the \emph{global controller}.
For multi-node jobs, the \emph{global controller} aggregates the statistics reported from all \emph{local controllers} where the job is being executed.
For instance, in Fig.~\ref{fig:padll-high-level-arch}, to observe the metrics of \emph{job}$_1$, the \emph{global controller} aggregates statistics from \emph{local controllers} of \emph{compute nodes 1} and \emph{2}.

\vspace*{.1\baselineskip}\noindent\emph{Compute:}
the \emph{global controller} then verifies if all policies are being met, by correlating the QoS limits defined by the system administrator (\emph{e.g.,} maximum metadata rate defined for \emph{job}$_1$) and the rates reported from stages (\emph{e.g.,} actual rate experienced by \emph{job}$_1$).
If the imposed limits are not being met, due to workload or system variations, it generates new rates (\emph{i.e.,} \emph{rules}) to the uncompliant stages. 

\vspace*{.1\baselineskip}\noindent\emph{Enforce:}
all generated rules are then submitted to \emph{local controllers}, which in turn will be submitted to the corresponding stages, where token-buckets will be adjusted with a new rate.

\vspace*{.1\baselineskip}\noindent\emph{Sleep:}
as a complementary step, \emph{sleep} defines the periodicity of control cycles (\emph{e.g.,} perform the aforementioned control steps at 1-second intervals).
With small intervals stages will be adjusted more frequently and impose higher control overhead (\emph{i.e.,} continuously collect statistics and enforce rules), while with larger intervals jobs may become unsupervised for long periods, which can be harmful under volatile workloads.

\paragraph{Orchestrating stages of distributed jobs}
Every time a single or multi-node job starts, its stages are initialized and connected to the corresponding \emph{local controllers}.
Each stage sends to the controller information that characterizes the job and the node it is running, such as the \emph{jobID}, \smalltexttt{PID}, \emph{hostname}, and \emph{userID}.
\emph{Local controllers} synchronize this information with the \emph{global controller}.
Based on this, the control plane knows which job each stage respects to, orchestrating the stages that belong to the same \smalltexttt{job-ID} as a single one.


\subsection{Implementation}
\label{subsec:implementation}

We have implemented \SYS's \emph{data} and \emph{control planes} with 16K and 6K lines of C++ code, respectively. 
All artifacts discussed in this paper are publicly available (\cref{sec:availability}).

\paragraph{Transparently intercepting POSIX calls}
The data plane uses \texttt{LD\_PRE\-LO\-AD} to transparently intercept POSIX calls and handle them before being submitted to the PFS.
It supports 42 system calls from different operation classes, including data, metadata, extended attributes, and directory management. 

\paragraph{Request differentiation and rate limiting}
The logic for differentiating and rate limiting requests (\emph{e.g.,} queues, token-buckets) was built using \PAIO~\cite{PAIO:2022:Macedo}, a framework for building custom-made, user-level storage data planes.

\paragraph{Communication}
Communication between controllers is established through RPC calls, using the gRPC framework~\cite{gRPC}, while communication between \emph{local controllers} and data plane stages is established using UNIX Domain Sockets.
For the latter, we adopt the control interface proposed in \PAIO.

\paragraph{Control delegation}
Currently, \emph{local controllers} act as proxies that aggregate statistics from stages before being dispatched to the \emph{global controller}, and forward enforcement rules to the respective stages. 
We defer the delegation of control logic (\emph{i.e.,} control partitioning) from the \emph{global controller} to \emph{local controllers} to future work.

\section{Control Algorithms}
\label{sec:control-algorihtms}
\vspace*{-5pt}

We now present state-of-the-art control algorithms supported by \SYS, and introduce a new QoS algorithm for preventing resource over-provisioning.
In \SYS, control algorithms can be either \emph{static} (\cref{subsec:static-alg}) or \emph{dynamic} (\cref{subsec:dynamic-alg}).
For all, we consider the amount of operations serviced by the PFS, either data (bandwidth) or metadata (IOPS), as the shared resource to be distributed among jobs.
Further, we define \emph{Max}$_R$ as the maximum throughput of either data or metadata that a given PFS can service.

\subsection{Static Control Algorithms}
\label{subsec:static-alg}

\emph{Static} control algorithms enable defining, for each job, \emph{fixed I/O limits} for accessing shared storage resources.

\paragraph{Uniform}
In a \emph{uniform rate distribution} jobs are throttled with a fixed limit throughout their execution, regardless of their size (\emph{i.e.,} number of compute nodes), duration, and workload (\emph{e.g.,} access pattern, I/O load, dataset size).
Such an approach is useful to equally distribute resource shares among jobs.

\paragraph{Priority}
In a \emph{priority-based rate distribution}, PFS resources are distributed based on a given priority, where jobs with higher priority have access to a larger resource share.

\vspace*{.2\baselineskip}
These algorithms follow a similar approach to those of \emph{cgroups blkio}~\cite{blkio} and OOOPS~\cite{OOOPS:2020:Huang}. 
Analogously, as I/O limits are fixed throughout the jobs' execution, these do not leverage from \SYS's global visibility, which can result in a misuse of system resources.
Specifically, jobs cannot be dynamically adjusted when (1) there are leftover resources (\emph{e.g.,} a job ended its execution and released its resource share), leading to \emph{under-provisioning}; or (2) jobs are assigned with shares larger than they need (\emph{e.g.,} job submits operations at a rate lower than the defined limit), experiencing \emph{over-provisioning}.

\subsection{Dynamic Control Algorithms}
\label{subsec:dynamic-alg}
\vspace*{-5pt}

\emph{Dynamic} control algorithms enable assigning, to each job, resource shares that change over time (based on \emph{soft} and \emph{hard I/O limits}), being adaptable to workload or system variations (\emph{e.g.,} jobs entering or leaving the system, volatile workloads). 
These algorithms are implemented in a \emph{feedback control loop} and are executed in the \emph{global controller}.

\paragraph{Proportional sharing}
To ensure I/O fairness while preventing under-provisioning scenarios, we implemented a max-min fair share control algorithm that enforces \emph{per-job rate reservations}, similar to those in \cite{TBF:2017:Qian,PAIO:2022:Macedo,IOFlow:2013:Thereska,Retro:2015:Mace}.
At any given time, jobs are allocated with I/O resources in order of increasing \emph{demands} (\emph{i.e.,} defined QoS limit), where (1) no job gets a share larger than its \emph{demand} and (2) jobs with unsatisfied demands get equal shares of resources.
Then, whenever there are leftover resources -- for example, the current metadata rate used by all jobs has not reached \emph{Max}$_R$ -- the algorithm distributes them across active jobs in a proportional manner.

While this algorithm is well-suited for workloads with sustained I/O load, it is suboptimal under volatile workloads.
Specifically, the algorithm always allocates a share of the I/O resources to jobs, regardless of their I/O load; if a given job follows a volatile workload, the algorithm may assign a share larger than it needs, resulting in over-provisioning. 
We refer to this behavior as \textbf{\emph{false resource allocation}}.

\begin{algorithm}[t]
    \footnotesize
    \begin{algorithmic}[1]

    \caption{\emph{Prop. sharing w/o false resource allocation}}
    \label{alg:maxmin-sus}

    \Require \emph{Max}$_{R}$ \emph{= N}; \emph{Active} $> 0$; \emph{demand}$_i > 0$; \emph{usage}$_i > 0$; $0 \leq$ {\normalsize$\varepsilon$} $\leq 1$

    \small
    \State \{\emph{usage}$_0$, $...$\ , \emph{usage}$_{Active-1}$\} $\gets$ \emph{collect\ ()}
    
    \State \emph{left}$_{R} \gets $\emph{Max}$_R$
    
    \For{$i=0$ in [0, \emph{Active}$ - 1$]}
        \State \emph{fair\_share} $\gets \frac{left_{R}}{Active - i}$

        \If{\emph{usage}$_i \leq$ \emph{demand}$_i$}
            \State \emph{threshold}$_i \gets ($\emph{demand}$_i\ -$ \emph{usage}$_i)\ *\ ${\normalsize$\varepsilon$}

            \State \emph{rate}$_i \gets min\ ($\emph{usage}$_i$ + \emph{threshold}$_i$, \emph{fair\_share}$)$
        \Else
            \State \emph{rate}$_i \gets min\ ($\emph{demand}$_i$, \emph{fair\_share})
        \EndIf

        \State \emph{left}$_{R} \gets$ \emph{left}$_{R}\ -$ \emph{rate}$_i$ 
    
    \EndFor

    \State \emph{total\_usage} $\gets \sum_{j=0}^{Active - 1} usage_{j}$

    \For{$i=0$ in [0, \emph{Active}$ - 1$]}
        \State \emph{usage\_proportion}$_i \gets \frac{usage_i}{total\_usage}$

        \State \emph{rate}$_i \gets$ \emph{rate}$_i$ + \emph{usage\_proportion}$_i\ *$ \emph{left}$_{R}$  
    \EndFor

    \State \emph{enforce} $(\{$\emph{rate}$_0$, $...$\ , \emph{rate}$_{Active-1}\})$

    \State \emph{sleep (loop\_interval)}

    \end{algorithmic}
\end{algorithm}

\paragraph{Proportional sharing without false allocation}
We propose a new proportional sharing algorithm that prevents false resource allocation to ensure storage QoS under volatile workloads (Alg.~\ref{alg:maxmin-sus}), entitled as \ALG.
Briefly, rather than assigning resource shares exclusively based on the number of active jobs in the system and their \emph{demands}, we consider the actual usage (\emph{i.e.,} I/O load) of each job and redistribute resources in a max-min fair share manner based on those observations.

In more detail, the algorithm performs the following steps.
First, it collects statistics from each active job's stage to determine its actual rate, given by \emph{usage}$_i$ (\smalltexttt{1}). 
For each active job, the algorithm computes its \emph{fair\_share} (\smalltexttt{4}) and verifies if the current rate (\emph{usage}$_i$) is lower than its \emph{demand} (\smalltexttt{5}).
Under this scenario, \emph{job}$_i$ can be serviced at a rate lower than its \emph{demand}.
As such, it assigns the minimum between \emph{fair\_share} and \emph{usage}$_i$~{+}~\emph{threshold}$_i$ (\smalltexttt{7}).
\emph{Threshold}$_i$ is computed based on the product of a configurable $\varepsilon$ value and the difference between \emph{demand}$_i$ and \emph{usage}$_i$, and is used to absorb the rate of highly volatile workloads (\smalltexttt{6}).
If \emph{usage}$_i$ is higher than \emph{demand}$_i$, the controller assigns the minimum between \emph{demand}$_i$ and the \emph{fair\_share} (\smalltexttt{8-9}).
The algorithm then distributes leftover rate (\emph{left}$_R$) across actives jobs (\smalltexttt{11-14}).
Specifically, it computes the overall rate used by all jobs (\smalltexttt{11}), and assigns \emph{left}$_R$ based on their usage proportion, given by \emph{usage\_proportion}$_i$ (\smalltexttt{13-14}).
Finally, the \emph{global controller} generates rules (\texttt{enforce}) to be submitted to each \emph{local controller} (\smalltexttt{15}), and sleeps for \emph{loop\_interval} before beginning a new control cycle (\smalltexttt{16}).

\vspace*{-5pt}
\section{Evaluation}
\label{sec:evaluation}
\vspace*{-7.5pt}

\begin{figure*}[t]
    \centering
    \includegraphics[width=1\linewidth]{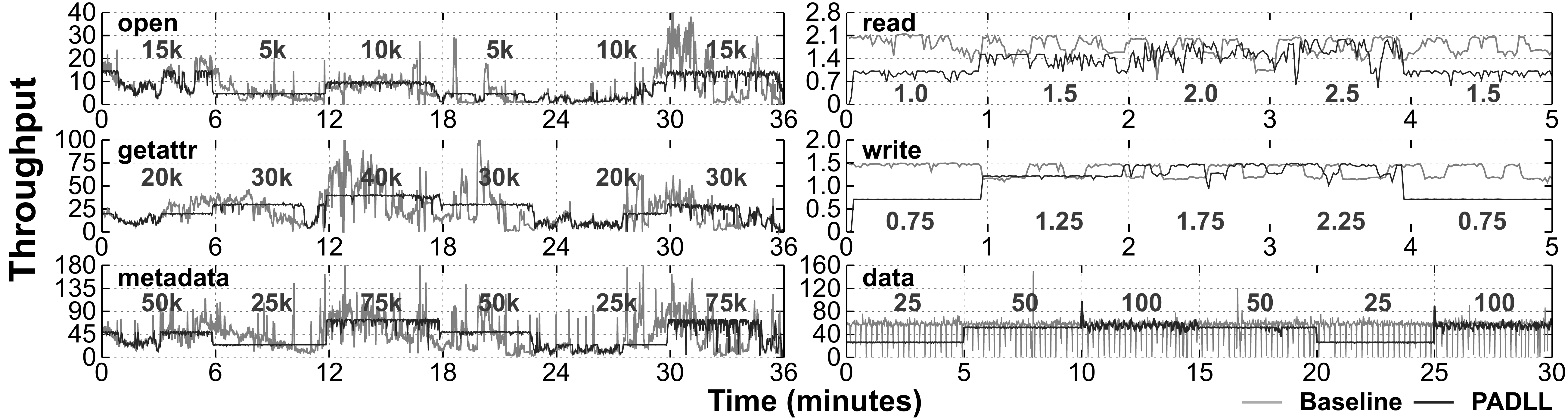}

    \caption{   
        \textbf{Per-operation type and class rate limiting.}
        Experiments show that \SYS can enforce different rate limits over different POSIX operations and granularities, including \texttt{open}, \texttt{getattr}, \texttt{read}, \texttt{write}, \texttt{metadata}, and \texttt{data}.
        Metadata operations are presented in IOPS (kops/s), while data operations in I/O bandwidth (namely, GiB/s for \texttt{read}/\texttt{write} and MiB/s for \texttt{data} experiments).
    } 
    
    \label{fig:operation-type-rate-limiting}
    \vspace*{-15pt}
\end{figure*}

\noindent
Our evaluation seeks to answer the following questions:

\begin{itemize}[nosep, leftmargin=*]
    \item Can \SYS control I/O workflows at different granularities?
    \item Can \SYS enforce QoS policies over concurrent jobs?
    \item What is the performance of \SYS control and data planes?
\end{itemize}

\paragraph{Experimental testbed}
Experiments were conducted on three hardware configurations.
\begin{itemize}[nosep, leftmargin=*]
\item \textbf{Configuration \emph{A}:} respects to compute nodes of the \abci supercomputer~\cite{abci}, equipped with two 20-core Intel Xeon processors, 384~GiB of RAM, and an InfiniBand EDR network card, running CentOS 7.5. 
The PFS is a \emph{dedicated} DDN ExaScaler Lustre composed of 2 MDSs in a hot-standby configuration, backed by 2 MDTs, and 24 OSTs that provide 359 TiB of storage capacity.
\item \textbf{Configuration \emph{B}:} respects to compute nodes of the \frontera supercomputer~\cite{Frontera:2020:Stanzione}, equipped with two 16-core Intel Xeon processors, 128 GiB of RAM, four NVIDIA Quadro RTX 5000 GPUs, and a Mellanox InfiniBand FDR network card, running CentOS 7.9.
The production PFS is a Lustre file system composed of 4 MDSs, each with a single MDT, and 32 OST nodes with 22 PiB of storage capacity.
\item \textbf{Configuration \emph{C}:} respects to compute nodes of the \frontera supercomputer, equipped with two 28-core Intel Xeon processors, 192 GiB of RAM, and a Mellanox InfiniBand HDR-100 network card, running CentOS 7.9.
The production PFS is the same as hardware configuration \testbedB.
\end{itemize}

\paragraph{Benchmarks and workloads}
We conducted experiments using both data and metadata workloads.
For data workloads we used IOR~\cite{IOR:2007:Shan}, which performs a \texttt{read}/\texttt{write} workload that sequentially reads/writes from/to a single file with 875 GiB using POSIX-compliant system calls. 
We also used TensorFlow~\cite{TensorFlow:2016:Abadi}, an AI framework used for DL training that is predominately executed in today's HPC clusters~\cite{DeepIO:2018:Zhu,RevisitingIOBehavior:2019:Patel}.

To generate realistic metadata workloads, we implemented a \emph{trace replayer} that submits metadata operations with an identical request distribution as the one observed from the logs collected at \abcifs (\cref{subsec:study}).
The \emph{replayer} is multi-threaded, and each thread submits operations at a rate that follows the same performance curve as original logs.
The rate of each operation was scaled-down to half, due to the difference in size between \abcifs and configuration \testbedA's PFS.
The execution period was accelerated, and each second of the \emph{replayer} corresponds to a minute's worth of operations in the original trace.

\paragraph{Methodology}
For all experiments, the \emph{global controller} runs at a dedicated compute node, while \emph{local controllers} execute co-located with each job instance and respective data plane stages. 
Metadata experiments were conducted with hardware configuration \testbedA (\cref{subsec:eval-operation-type}--\cref{subsec:eval-job}), and data experiments with configurations \testbedB (\cref{subsec:eval-operation-type}, \cref{subsec:eval-operation-class}) and \testbedC (\cref{subsec:scalability}). 
All experiments were conducted using the shared PFS.
Across all testing scenarios, two setups were used:
\baseline, which represents the benchmark without using \SYS, and 
\padll, where POSIX operations submitted by the benchmark are intercepted by \SYS and throttled at a given rate.


\subsection{Per-operation type rate limiting}
\label{subsec:eval-operation-type}

We first demonstrate how \SYS enables system administrators to control the rate of specific operations.
Fig.~\ref{fig:operation-type-rate-limiting} depicts the results of all setups under different operation types. 

\paragraph{Workload configuration}
Both \emph{trace replayer} and \emph{IOR} were configured to submit a single operation type -- namely, \texttt{open} and \texttt{getattr} (\emph{left}), and \texttt{read} and \texttt{write} (\emph{right}).

\paragraph{\SYS configuration}
For all experiments, \SYS was configured to throttle operations with a static rate, whose value changes every $N$~minutes (6~minutes for metadata and 1~minute for data operations) upon instruction of the system administrator (\emph{i.e.,} rule defined on the control plane).

\paragraph{Results}
At all times, \padll is able to control the rate of all operations, never exceeding the configured limits.
Over several periods, \padll follows the same performance curve as \baseline, as observed in \texttt{open} between 23 and 29 minutes (\emph{i.e.,} periods where the black line is not flat).
This is because, the limit set by the system administrator (for that interval) is higher than the operations submitted by the \emph{replayer}.
Analogously, we also observe periods where \padll achieves higher throughput than \baseline, as observed in \texttt{getattr} between 8 and 12 minutes. 
This happens when operations are being aggressively rate limited (\emph{i.e.,} the original rate is higher than the defined limit), creating a backlog of operations to be executed later when there are enough available resources.

We observe similar results for data-oriented operations, namely \texttt{read} and \texttt{write}. 
However, contrarily to configuration \testbedA where requests are submitted to a dedicated PFS, we observe more variability, as experiments were conducted over a file system shared with multiple concurrent jobs.


\subsection{Per-operation class rate limiting}
\label{subsec:eval-operation-class}

We now demonstrate how \SYS controls the workflows of a given operation class, namely \texttt{metadata} and \texttt{data}.

\paragraph{Workload configuration}
We followed a similar methodology as in~\cref{subsec:eval-operation-type}.
For \texttt{metadata}, operations are submitted from a multi-node job made of 4 \emph{trace replayer} instances, each running on a dedicated compute node.
Each instance spawns 8 threads that submit different types of metadata operations (same as those discussed in \cref{subsec:study}).
To demonstrate \SYS's general applicability, we conducted the \texttt{data} experiments using a distributed TensorFlow job over 4 compute nodes, running version 2.3.2 with the LeNet training model~\cite{LeNet:1998:LeCun} and configured with a batch size of 128 TFRecords.
We used the ImageNet dataset ($\approx$150 GiB)~\cite{Imagenet:2015:Russakovsky}, hosted at the shared PFS.

\paragraph{\SYS configuration}
Similar to \cref{subsec:eval-operation-type}, \SYS was configured to throttle operations with a static rate, whose value changes every 6 minutes for metadata workloads and 5 minutes for data workloads.
Fig.~\ref{fig:operation-type-rate-limiting} (\emph{bottom}) depicts the obtained results.
The throughput corresponds to the accumulated rate of all \emph{replayer} instances (in IOPS) or TensorFlow workers (in I/O bandwidth).

\paragraph{Results}
In both \texttt{data} and \texttt{metadata} experiments, \padll effectively controls the rate of all workflows throughout the overall observation.
In several periods, \padll matches or achieves higher throughput performance than \baseline; we draw similar observations as in \cref{subsec:eval-operation-type}.
Complementary, these experiments also demonstrate that \SYS can achieve QoS limits even for distributed jobs.


\subsection{Per-job QoS control}
\label{subsec:eval-job}

We now demonstrate how \SYS achieves per-job QoS control in HPC storage systems by orchestrating the metadata workflows of all active jobs. 
Under this scenario, metadata operations are treated as a finite and shared I/O resource, and for the PFS to provide sustained I/O performance, jobs need to meet specific metadata service level objectives (SLOs).

\paragraph{Workload configuration}
At all times, there are at most four jobs in the system, each running 4 \emph{trace replayer} instances in dedicated compute nodes (16 in total) and submitting metadata operations. 
Jobs are incrementally added to the system every 3 minutes. 
We consider that the system administrator defines a maximum rate of metadata operations (\emph{Max}$_R$) that can be submitted to the targeted PFS, being set at 110~kops/s (red dashed lines).
The trace used in the experiments corresponds to the metadata operations of all MDT servers of \abcifs.

\begin{table}[t]
    \centering
    \footnotesize
    \caption{Per-Job QoS control testing scenarios.}
    \vspace*{-5pt}
    \begin{tabular}{rccc}
        \toprule
                     & \textbf{Test. scenario \#1}  & \textbf{Test. scenario \#2}  & \textbf{Test. scenario \#3}  \\ \midrule
        \emph{\textbf{Job$_1$}} & 25\% -- 15~kops/s & 15\% -- 15~kops/s & 15\% -- 40~kops/s \\ 
        \emph{\textbf{Job$_2$}} & 25\% -- 25~kops/s & 20\% -- 25~kops/s & 20\% -- 25~kops/s \\ 
        \emph{\textbf{Job$_3$}} & 25\% -- 30~kops/s & 20\% -- 30~kops/s & 20\% -- 30~kops/s \\ 
        \emph{\textbf{Job$_4$}} & 25\% -- 40~kops/s & 45\% -- 40~kops/s & 45\% -- 15~kops/s \\ \bottomrule
    \end{tabular}
    \label{table:testing-scenarios}
\end{table}

\paragraph{Testing scenarios}
To provide a comprehensive evaluation testbed, we consider three testing scenarios with varying load proportions and rate limits. 
Table~\ref{table:testing-scenarios} depicts the load proportion and the metadata rate limit for each combination of testing scenario and job.
\emph{Testing scenario \#1:} jobs follow the same workload but are assigned with different priorities.
\emph{Testing scenario \#2:} jobs have different load proportions and rate limits are assigned proportionally to each job's load (\emph{i.e.,} jobs with lower load are assigned with lower priority).
\emph{Testing scenario \#3:} jobs follow the same load proportions as \emph{testing scenario \#2}, but the rate limits of jobs 1 and 4 are switched.

\paragraph{Setups}
Experiments were conducted under five setups.
\emph{Baseline} represents the current setup supported at most supercomputers, where jobs execute without any throttling.
The remainder setups are rate limited with \SYS, with a maximum combined rate of \emph{Max}$_R$, and respect to the control algorithms discussed in \cref{sec:control-algorihtms}.
In \emph{Uniform}, each job is rate limited to 27.5~kops/s, while \emph{Priority}, \emph{Proportional sharing (PSharing)}, and \emph{\ALG}, jobs are assigned different rates, as depicted in Table~\ref{table:testing-scenarios}.
For \emph{PSharing}, these limits represent the per-job maximum rate when all jobs are active, while for \emph{\ALG} represent the per-job maximum rate when all jobs are active and each job's \emph{usage} is higher than its \emph{demand}.


\begin{figure}[t]
    \centering
    \includegraphics[width=1\linewidth]{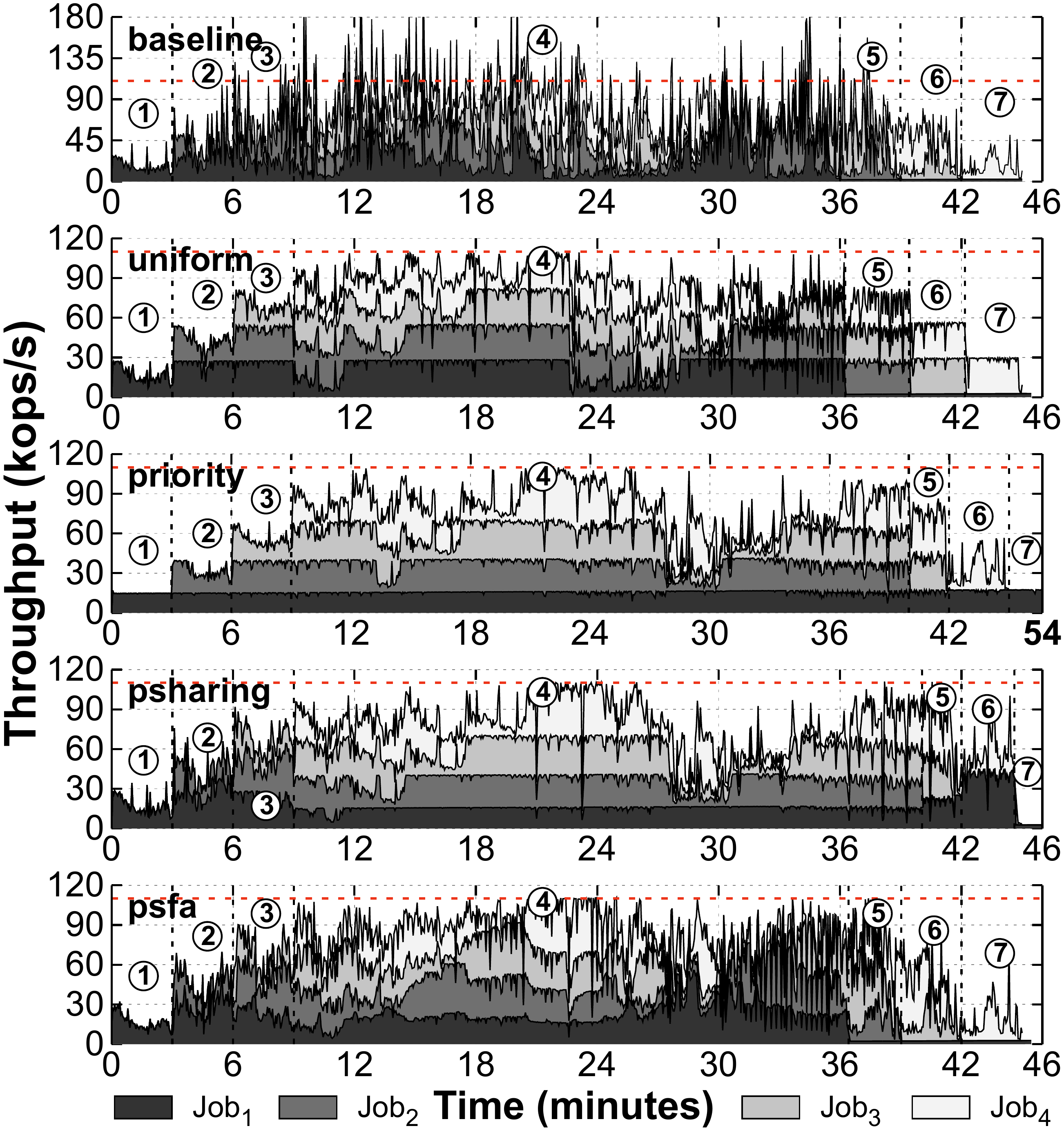}
    \caption{   
        \textbf{Per-job metadata control} over Baseline, Uniform, Priority, PSharing, and \ALG setups under \textbf{testing scenario \#1}.
    }
     \label{fig:testing-scenario-1}
\end{figure}

\paragraph{Testing scenario \#1}
Fig.~\ref{fig:testing-scenario-1} depicts, for each setup, the metadata rate of all jobs at 1-second intervals under \emph{testing scenario \#1}. 
Experiments include seven phases (\numlightcirclesans{1}--\numlightcirclesans{7}), each marking when a given job enters or leaves the system.

\emphparagraph{Baseline}
Experiments were executed over 45 minutes.
Each job executes over 36 minutes and leaves the system in the same order as it entered.
Throughout the entire execution, we observe that the workload is extremely volatile and bursty, with peaks that reach 300~kops/s and several periods where the file system continuously serves requests over \emph{Max}$_R$.

\emphparagraph{Uniform}
Experiments were executed over 45 minutes.
Whenever a new job is added, it is provisioned with its assigned rate (27.5~kops/s).
\SYS ensures, to all jobs, sustained metadata throughput and eliminates existing burstiness. 
However, while this setup is useful to equally distribute metadata rate among jobs, it does not allow them to execute with different priorities. 

\emphparagraph{Priority}
Experiments were executed over 54 minutes.
Similarly to \emph{Uniform}, \SYS ensures that all jobs are provisioned with their assigned rate.
However, when a job is set with low priority, its execution may take longer than its corresponding unthrottled version since metadata operations are rate limited more aggressively.
We observe this in \emph{Job}$_1$, where its execution takes 18~minutes longer than in previous setups, and occurs because the algorithm does not leverage from leftover metadata rate (\emph{i.e.,} \numlightcirclesans{1}-\numlightcirclesans{3} and \numlightcirclesans{5}-\numlightcirclesans{7}), as discussed in \cref{subsec:static-alg}.

\emphparagraph{PSharing} 
Experiments were executed over 45 minutes. Whe\-ne\-ver a new job enters (\numlightcirclesans{1}-\numlightcirclesans{3}) or leaves the system (\numlightcirclesans{5}-\numlightcirclesans{7}), it is assigned with its proportional share.
When all jobs are running (\numlightcirclesans{4}), they are assigned with their demanded rate.
Compared to \emph{Priority}, \emph{PSharing} distributes leftover metadata rate, which enables improving \emph{Job}$_1$'s performance by 9~minutes.
However, since the workload is volatile and bursty, we observe several periods with \emph{false resource allocation} (\cref{subsec:dynamic-alg}), being particularly noticeable in the 12--20 minutes and 25--37 minutes intervals.
During these periods one or more jobs are over-pro\-vi\-sio\-ned, and the exceeding resources could be used to improve the performance of the remainder jobs (\emph{e.g.,} \emph{Job}$_1$ still took 9~minutes longer to execute than in \emph{Baseline}).

\emphparagraph{\ALG}
Experiments were executed over 45 minutes.
All jobs complete their execution in the same time as their corresponding unthrottled versions.
Throughout the overall execution, the \emph{\ALG} algorithm continuously adjusts the limit of each job based on the actual rate it is using, preventing over-provisioning. 
Specifically, in the 12--20 minutes period, \emph{\ALG} assigns unused metadata rate to \emph{Job}$_1$, \emph{Job}$_2$, and \emph{Job}$_3$, temporarily having more rate than their \emph{demand}.
The same is observed for \emph{Job}$_1$ and \emph{Job}$_2$ in the 25--37 minutes period.

\begin{figure}[t]
    \centering
    \includegraphics[width=1\linewidth]{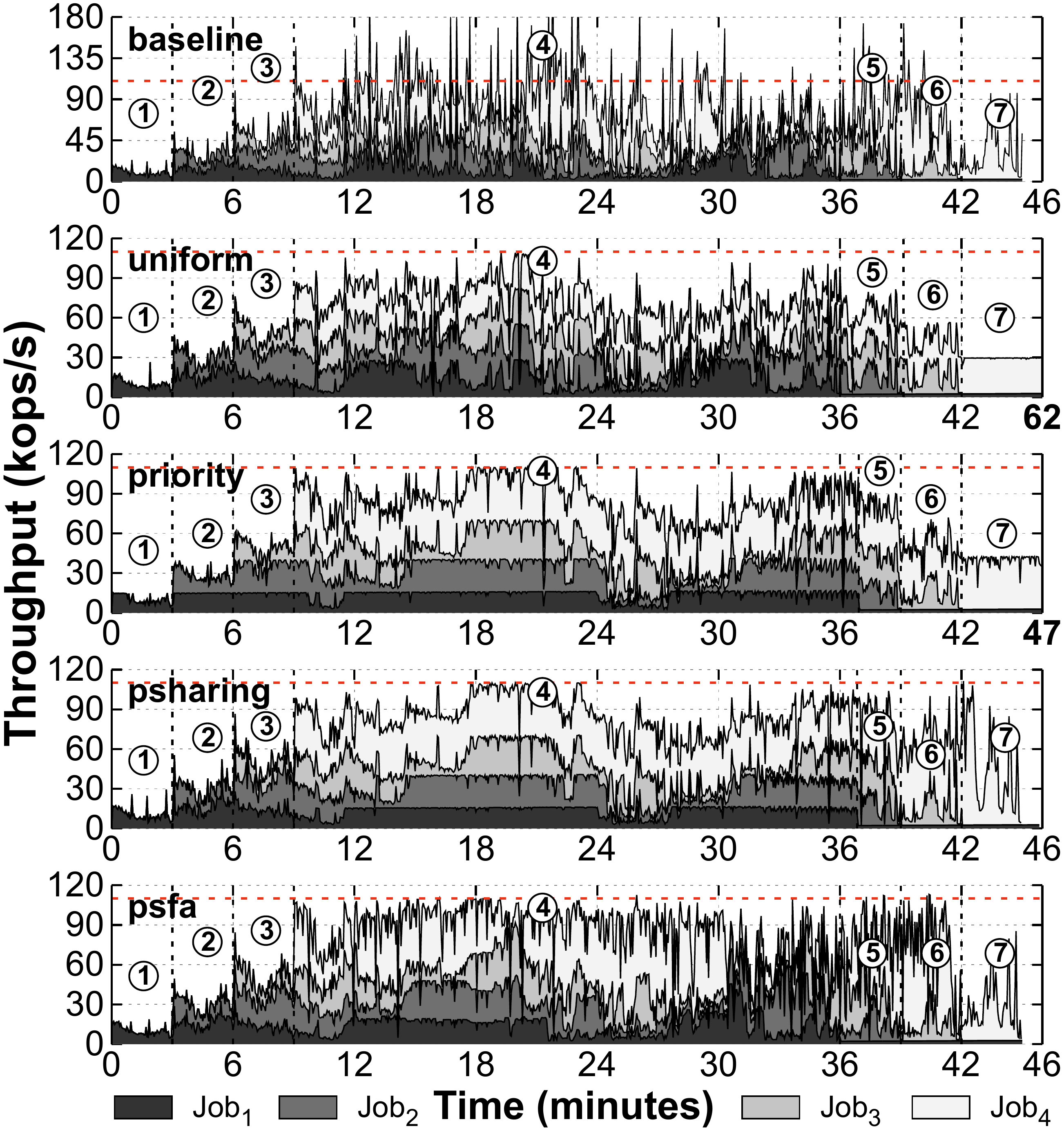}
    \caption{   
        \textbf{Per-job metadata control} over Baseline, Uniform, Priority, PSharing, and \ALG setups under \textbf{testing scenario \#2}.
    }
     \label{fig:testing-scenario-2}
\end{figure}

\paragraph{Testing scenario \#2}
Fig.~\ref{fig:testing-scenario-2} depicts, for each setup, the metadata rate of all jobs at 1-second intervals under \emph{testing scenario \#2}.

\emphparagraph{Baseline}
Experiments executed over 45 minutes.
As the overall metadata load is the same across all testing scenarios, we observe similar volatility and burstiness as in \emph{testing scenario \#1}.
The key difference is that now \emph{Job}$_4$ generates a major part of the metadata load, while \emph{Job}$_1$ has significantly lower load.

\emphparagraph{Uniform}
Experiments executed over 62 minutes.
Similarly to \emph{testing scenario \#1}, whenever a new job is added it is provisioned with its static rate (27.5~kops/s). 
However, as \emph{Job}$_4$ now generates 45\% of the overall metadata load, \SYS aggressively rate limits it, resulting in a longer execution period (\emph{i.e.,} 17~minutes longer than \emph{Baseline}).
On the other hand, \emph{Job}$_1$ experiences over-provisioning for most of its execution, given that it only generates 15\% of the overall workload.

\emphparagraph{Priority}
Experiments executed over 47 minutes.
Due to the decreased metadata load, \emph{Job}$_1$ finishes approximately at the same time as in \emph{Baseline}.
On the other hand, \emph{Job}$_4$ takes 2~minutes longer to complete.
Specifically, even though \emph{Job}$_4$ is set with a larger resource share (40~kops/s), due to its large metadata load, \SYS aggressively rate limits it throughout the overall execution period, resulting in a large backlog of metadata operations to be performed, as observed in \numlightcirclesans{7}.

\emphparagraph{PSharing}
Experiments executed over 45 minutes.
Contrarily to static setups, since \emph{PSharing} distributes leftover metadata rate whenever it is available, jobs complete their execution in 36~minutes (as in \emph{Baseline}).
However, similarly to the observations made in \emph{testing scenario \#1}, this setup experiences periods with \emph{false resource allocation}, being especially noticeable in the 12--18, 23--35, and 36--42 minutes intervals.

\emphparagraph{\ALG}
Experiments executed over 45 minutes.
\emph{\ALG} maximizes the use of metadata resources by reallocating unused rate from over-provisioned jobs.
For instance, during the 23--30 minutes period, Job$_4$ improves its performance by leveraging from unused metadata rate of the other jobs. 
Interestingly, during the 31--36 interval, \emph{\ALG} demonstrates lower usage of resources compared to \emph{Priority} and \emph{PSharing}.
This is because up to the 31-min mark, \emph{\ALG} allocates enough rate to all jobs (either from leftovers or over-provisioning) that allowed them to conduct any accumulated backlog of metadata operations. 


\begin{figure}[t]
    \centering
    \includegraphics[width=1\linewidth]{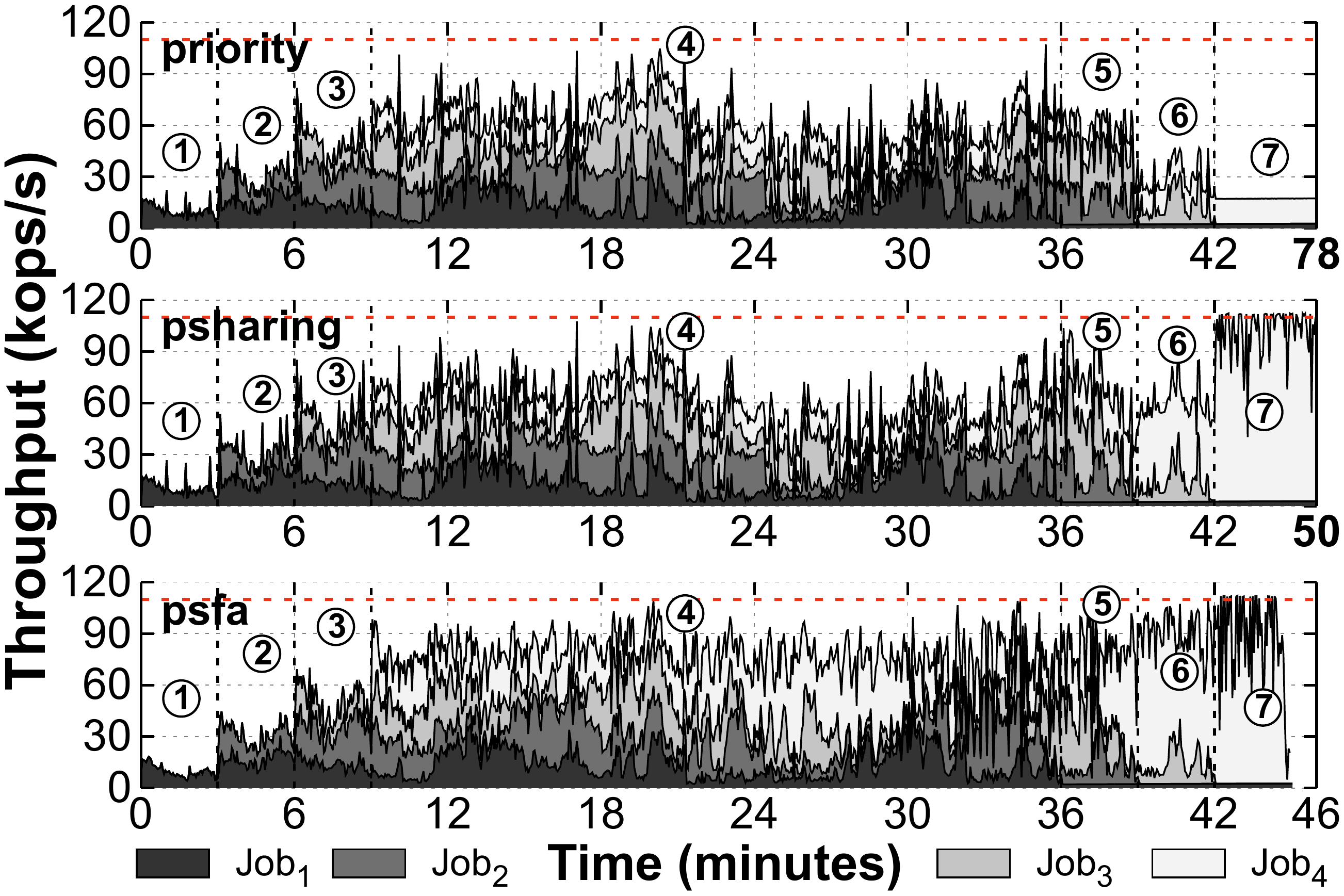}
    \caption{   
        \textbf{Per-job metadata control} over Priority, PSharing, and \ALG setups under \textbf{testing scenario \#3}.
    }
     \label{fig:testing-scenario-3}
\end{figure}

\paragraph{Testing scenario \#3}
Fig.~\ref{fig:testing-scenario-3} depicts, for each setup, the metadata rate of all jobs at 1-second intervals under \emph{testing scenario \#3}. 
Experiments conducted for \emph{Baseline} and \emph{Uniform} setups are the same as in \emph{testing scenario \#2}, as both metadata load and rate limits remain unchanged. 
We draw identical observations.

\emphparagraph{Priority}
Experiments executed over 78~minutes.
As a result of being assigned with the lowest priority while having the highest load, \emph{Job}$_4$ takes 33~minutes longer to complete its execution.
Noticeably, during \numlightcirclesans{7}, \SYS aggressively rate limits operations at a constant rate of 15~kops/s, not leveraging from the remainder 95~kops/s available in the system.

\emphparagraph{PSharing}
Experiments executed over 50~minutes.
Since \emph{Job}$_4$ is the last to enter the system (\numlightcirclesans{4}), it only leverages from leftover rate when the other jobs complete their execution (\numlightcirclesans{5}--\numlightcirclesans{7}).
Thus, due to accumulated backlog (\numlightcirclesans{7}), the job submits metadata operations at a constant rate of \emph{Max}$_R$, being 28~minutes faster than \emph{Priority} but still requiring an additional 5~minutes to complete when compared to \emph{Baseline}.
On the other hand, \emph{Job}$_1$ is over-provisioned for most of its execution. 

\emphparagraph{\ALG}
Experiments executed over 45~minutes.
Since \emph{\ALG} prevents \emph{false resource sharing}, during the 12--42 minutes interval, a large share of unused metadata rate is assigned to \emph{Job}$_4$, which enables executing all accumulated backlog just under the 45-min mark (\numlightcirclesans{7}).
Note that \emph{\ALG} is able to reassign unused resources without compromising other policies; for instance, \emph{Job}$_1$ demonstrates the same performance curve as in setups with more strict policies, namely \emph{Priority} and \emph{PSharing}.


\paragraph{Summary}
Results demonstrate that \SYS can be used to enforce different QoS policies (through different control algorithms) over distributed metadata-aggressive jobs without exceeding \emph{Max}$_R$.
\emph{Uniform} is suited for scenarios where jobs have similar I/O load ({\#1}), while \emph{Priority} is appropriate for assigning larger metadata shares to jobs with higher load ({\#2}). 
\emph{PSharing} prevents under-provisioning, improving job execution time and overall resource usage by allocating leftover metadata rate ({\#1}, {\#2}). 
\emph{\ALG} maximizes the use of metadata rate, accelerating the performance of resource-hungry jobs without degrading over-provisioned ones ({\#1} -- {\#3}).

\subsection{\SYS performance, resource usage, and overhead}
\label{subsec:scalability}

\begin{figure}[t]
    \centering
    \includegraphics[width=0.80\linewidth]{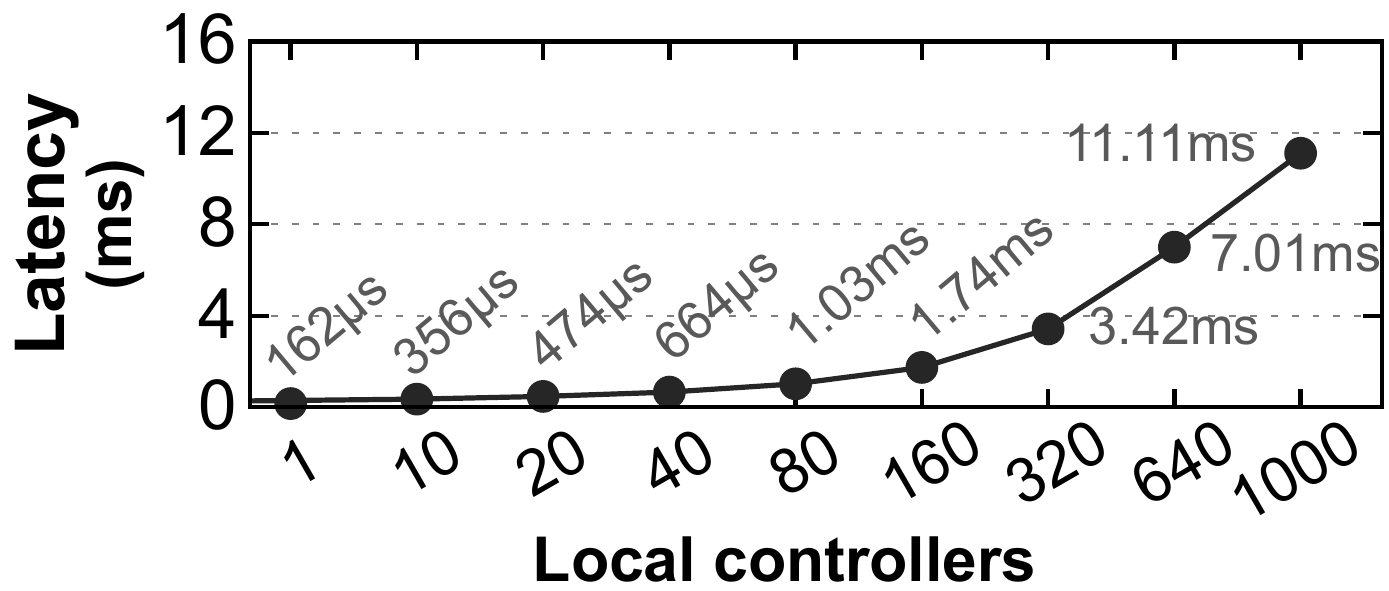}
    \vspace*{-5pt}
    \caption{\textbf{Average latency of control} cycles (in global controller) when the number of local controllers increases.}
    \label{fig:padll-latency}
\end{figure}

Finally, we demonstrate the performance of \SYS control and data planes, their impact on computational resources, and evaluate the overhead imposed over applications.

\paragraph{Control plane}
We conducted a set of experiments where the \emph{global controller} performs 250k iterations of \emph{\ALG} control algorithm's main phases, namely \emph{collect}, \emph{compute}, and \emph{enforce} (\cref{subsec:control-plane}).
In particular, the \emph{collect} and \emph{enforce} phases involve traversing the network multiples times per iteration.
The number of \emph{local controllers} submitting metrics and receiving enforcement rules is increased from 1 to 1,000.

Fig.~\ref{fig:padll-latency} depicts the obtained results, reporting the average latency of each control iteration.
Results demonstrate that latency increases with the number of connected controllers.
Up to 40 \emph{local controllers}, \SYS performs at $\mu s$-scale ranging between 162$\mu s$ and 664$\mu s$.
After that mark, the average latency per control cycle ranges between 1.03ms and 11.11ms.

These results show that the \emph{global controller} can orchestrate the overall system at \emph{ms}-scale, which fits the requirements of production workloads where these control cycles are tuned with larger time frames, as otherwise stages would be adjusted for every minimal workload change in the system. 
For instance, the experiments discussed in \cref{subsec:eval-job} were conducted with control cycles of 1-second intervals.
When reaching 1,000 nodes, we start to notice a higher utilization of network resources at the \emph{global controller}, and an increase in response time. 
This highlights the need to further research the scalability of this component as future work.


\begin{figure}[t]
    \centering
    \includegraphics[width=0.90\linewidth]{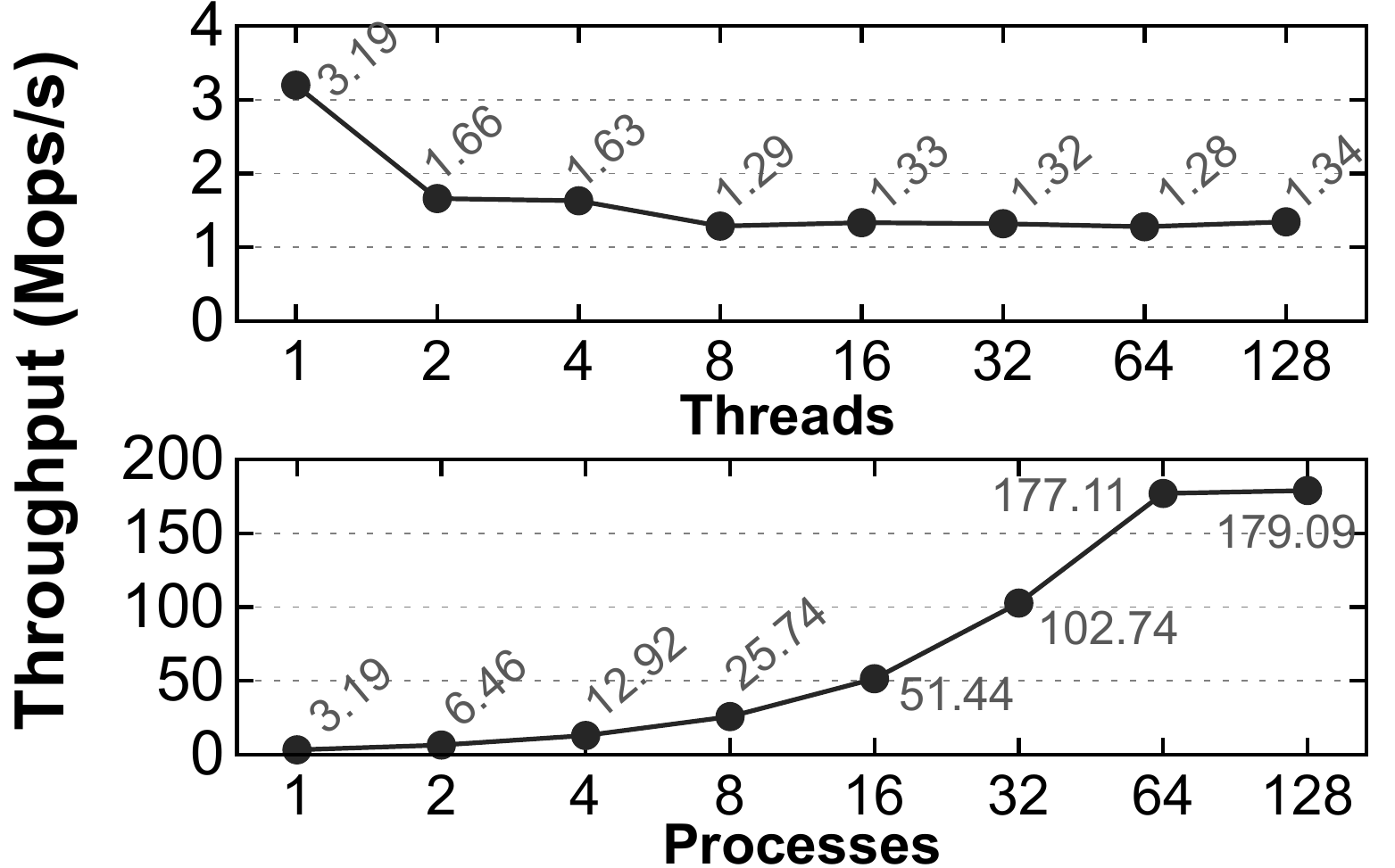}
    \vspace*{-5pt}
    \caption{\textbf{Data plane throughput} under a single stage that handles multi-threaded requests (top) and multiple stages that handle single-treaded requests (bottom).}
    \label{fig:padll-throughput}
\end{figure}

\paragraph{Data plane}
To evaluate the maximum performance achievable with the data plane, we implemented a benchmark that submits POSIX requests in a closed loop. 
Experiments were conducted under two scenarios: a \emph{single stage} that handles requests (\emph{i.e.,} \texttt{open}) under a varying number of threads (1--128); and \emph{multiple stages} (1--128), each handling single-threaded benchmark requests. 
Each thread submits 100M requests.
Stages are configured with the same number of \emph{channels} as client threads, and token-buckets are set with a rate large enough to not perform any throttling.
Requests follow the same path as discussed in \cref{subsec:data-plane}. 
Fig.~\ref{fig:padll-throughput} depicts obtained results for both \emph{single} (\emph{top}) and \emph{multi-stage} (\emph{bottom}) scenarios.

\noindent\emph{Single stage:}
under a single thread the stage handles requests at 3.19 Mops/s, while under multi-threaded settings (2--128 threads), due to internal locking and housekeeping management (\emph{i.e.,} register/access file descriptors (\cref{subsec:data-plane})), the performance ranges between 1.66 and 1.34 Mops/s.

\noindent\emph{Multiple stages:}
under this scenario, each stage handles the requests of a single benchmark instance (shared-nothing). 
As such, \SYS is able to handle requests up to 179.09 Mops/s, a performance increase of 133$\times$ compared to the previous scenario (under 128 threads).
Since the server is configured with 56 physical cores (hardware configuration \testbedC), the performance scales linearly up to 32 stages, ranging between 3.19 and 102.74 Mops/s. 
When configured with 64 and 128 stages, it achieves 177.11 Mops/s and 179.09 Mops/s, respectively.


\paragraph{Overhead}
To evaluate the overhead imposed by \SYS, we conducted experiments with a \pass setup, which respects to a scenario where POSIX operations submitted by the benchmark are handled by \SYS but are not rate limited.
We repeated \cref{subsec:eval-operation-type} and \cref{subsec:eval-operation-class} workloads.
When compared to \baseline, the overhead is negligible, never degrading performance more than 0.9\% across all experiments.

\paragraph{Resource usage}
We now discuss the resource usage impact imposed by \SYS.
In terms of \emph{network bandwidth}, the payload of the messages exchanged in each control cycle between the \emph{global} and each \emph{local controller} is approximately 200 bytes, which is negligible compared to the capacity of modern network devices. 
As for \emph{network latency}, as referred in the \emph{control plane performance} discussion (Fig.~\ref{fig:padll-latency}), \SYS is able to manage multiple \emph{local controllers} (and corresponding data plane stages) at $\mu s$-scale.
Regarding \emph{CPU} and \emph{memory} usage, the components that run co-located with the targeted jobs (\emph{i.e.,} data plane stages and local controller) impose minimal overhead, only increasing CPU by at most 5\% and memory by $\approx$100~MiB.
As such, \SYS imposes minimal overhead to all targeted jobs, being suited for the computational requirements of modern I/O infrastructures.

\section{Related Work}
\label{sec:related-work}

\paragraph{HPC storage QoS}
Many works, such as GIFT~\cite{GIFT:2020:Patel}, CALCioM~\cite{CALCioM:2014:Dorier}, IOrchestrator~\cite{IOrchestrator:2010:Zhang}, UShape~\cite{UShape:2011:Zhang}, and \emph{Gainaru et al}.~\cite{SchedulingUnderCongestion:2015:Gainaru}, are designed to mitigate I/O contention in HPC storage stacks but ignore the impact that metadata workflows have over the overall system performance.
\SYS is able to control the rate of both data and metadata workflows.
Other systems are directly implemented within core layers of the HPC I/O stack, including the PFS \cite{TBF:2017:Qian,SIREN:2018:Karki,SDQoS:2019:Hua,UShape:2011:Zhang,IOrchestrator:2010:Zhang}, scheduler~\cite{SchedulingUnderCongestion:2015:Gainaru}, and I/O libraries \cite{CALCioM:2014:Dorier,MappingScheduling:2020:Carretero}.
These solutions are intrusive and offer limited maintainability and portability.
\SYS actuates at the compute node level and does not require any changes to core layers of the HPC I/O stack.

Similarly to \SYS, OOOPS transparently intercepts and rate limits POSIX requests at compute nodes~\cite{OOOPS:2020:Huang}. 
However, it does not provide global visibility, being only capable of enforcing static policies that remain unchanged throughout the job execution.
\SYS on the other hand, enforces dynamic and cluster-wide QoS policies that require global visibility.

\paragraph{SDS systems}
\SYS builds on a large body of work on SDS~\cite{SDSsurvey:2020:Macedo}.
Systems like IOFlow, sRoute, and PSLO, actuate at the virtualization and block device layers, only controlling the rate of \texttt{read} and \texttt{write} requests~\cite{IOFlow:2013:Thereska,sRoute:2016:Stefanovici,PSLO:2016:Li,DifferentiatedServices:2011:Mesnier}.
Others, like Retro and Crystal, enforce resource management policies over distributed storage systems, but are directly implemented within the storage system itself, offering limited maintainability and portability~\cite{Retro:2015:Mace,Crystal:2017:Gracia}. 
SIREN enforces bandwidth policies over HPC storage systems, but is directly implemented within I/O forwarding and OSS nodes of OrangeFS~\cite{SIREN:2018:Karki,PVFS:2000:Carns}.
\SYS is a bare-metal solution that actuates at the compute node level and transparently intercepts and enforces POSIX requests, both data and metadata, before being submitted to the PFS. 
This makes it applicable over different applications and compatible with POSIX-compliant storage systems.

\paragraph{I/O optimizations}
Many works propose I/O optimizations to reduce the amount of operations submitted to the PFS by resorting to storage tiering and node-local storage~\cite{Monarch:2022:Dantas,Hermes:2018:Kougkas,UnifyFS:2017:Moody,Storage-Heterogeneity-Awareness:2022:Elshazly}, remote burst buffers~\cite{CARS:2019:Liang, Harmonia:2018:Kougkas,Storage-Heterogeneity-Awareness:2022:Elshazly}, data reduction techniques~\cite{HpcDeduplication:2012:Meister}, and optimized data formats~\cite{HDF5:2011:Folk,ADIOS:2008:Lofstead}, or improve the metadata management of large-scale file systems~\cite{SoMeta:2017:Tang,GraphMeta:2016:Dai,Brindexer:2020:Paul}.
While these can reduce the I/O pressure imposed over the PFS, they can still expose it to burstiness and unfairness, since I/O workflows are not rate limited.
On the other hand, \SYS rate limits all workflows destined towards the PFS.
Further, contrary to \SYS, several of these works are also intrusive to core layers of the HPC I/O stacks~\cite{CARS:2019:Liang, Harmonia:2018:Kougkas,SoMeta:2017:Tang,GraphMeta:2016:Dai,Brindexer:2020:Paul}.
While complementary to our work, these can be combined with \SYS to further improve the control of I/O workflows in HPC clusters.

\section{Conclusion}
\label{sec:conclusion}

We have presented \SYS, a storage middleware that enables QoS control of data and metadata workflows in HPC storage systems.
\SYS does not require changing any core layer of the HPC stack, and enforces storage policies with fine-granularity and global system visibility.
Results demonstrate that \SYS can enforce complex storage policies over concurrent metadata-aggressive jobs in holistic fashion, achieving I/O fairness, prioritization, and performance isolation.

With \SYS, we aim at supporting system administrators, researchers, and practitioners to effectively provide QoS control over large-scale HPC systems, and assist bridging the convergence of cloud and HPC infrastructures. 
\section*{Acknowledgements}

We thank AIST for providing access to computational resources of ABCI. 
We thank Cláudia Brito and Tânia Esteves for reviewing initial versions of this work.
This work is financed by: the ERDF - European Regional Development Fund, through the Operational Programme for Competitiveness and Internationalisation - COMPETE 2020 Programme under the Portugal 2020 Partnership Agreement, and by  National Funds through the FCT - Portuguese Foundation for Science and Technology, I.P. on the scope of the UT Austin Portugal Program within project BigHPC, with reference POCI-01-0247-FEDER-045924; through PhD Fellowships SFRH/BD/146059/2019 and PD/BD/151403/2021; and the UT Austin Portugal Program, a collaboration between the Portuguese Foundation of Science and Technology and the University of Texas at Austin, award UTA18-001217.

\bibliographystyle{plain}
\bibliography{biblio}

\section*{Artifact Description/Evaluation}
\label{annex}

The paper proposes a new storage middleware that enables system administrators to proactively and holistically control the rate of data and metadata workflows to achieve QoS in HPC storage systems.
It is organized in two main components: the \emph{data plane} (\SYS) and the \emph{control plane} (\chef).

\paragraph{\SYS}
The data plane, named \SYS, is a multi-stage component that provides the building blocks for differentiating and rate limiting POSIX requests that are destined towards a given file system (\emph{e.g.,} Lustre, \texttt{ext4}).
Stages actuate at the compute node level, and use \texttt{LD\_PRELOAD} to transparently intercept POSIX requests (\emph{i.e.,} \texttt{libc.so} calls) from a given application and handling them before being submitted to the file system.
\SYS is written in C++17 and is publicly available at the \texttt{dsrhaslab/padll}\footnote{\SYS: \padllrepo} GitHub repository under a \emph{BSD 3-Clause} license.
The logic for differentiating and rate limiting requests was built using the \PAIO data plane framework~\cite{PAIO:2022:Macedo}.\footnote{\PAIO: \paiorepo}
Communication between \SYS stages and the control plane (\emph{local controllers}) is established using \emph{UNIX Domain Sockets}.

\paragraph{\chef}
The control plane, named \chef, is a logically centralized component with system-wide visibility that defines how all I/O requests should be handled.
It follows a hierarchical distribution, made of \emph{global} and \emph{local controllers}.
Local controllers are placed at compute nodes and manage all \SYS stages that are executing there.
The global controller executes at a dedicated compute node, and communicates with all local controllers with RPC calls through the gRPC framework.
Moreover, \chef is written in C++17 and is publicly available at the \texttt{dsrhaslab/cheferd}\footnote{\chef: \cheferdrepo} GitHub repository under a \emph{BSD 3-Clause} license.

\paragraph{Artifacts}
Both \SYS and \chef, alongside a \emph{trace replayer} and several scripts to reproduce the experiments of the paper are publicly available at Zenodo's artifact repository (\zenodo)~\cite{Padll:2023:Macedo}.

\begin{itemize}[nosep, leftmargin=*]
  \item \SYS and \chef have the same version as their corresponding publicly available GitHub repositories.
  \item The \emph{trace replayer} (\texttt{mdreplayer}) is used to generate realistic metadata workloads.
  It is multi-threaded, and each thread submits operations at a variable rate that is defined by the trace distribution.
  \item The \emph{scripts} folder includes bash scripts for installing and deploying all systems, as well as scripts to (1) reproduce experiments in the paper (\cref{subsec:eval-operation-type}--\cref{subsec:eval-job}) at the Frontera supercomputer, or (2) test a subset of them in a commodity setup (namely, \cref{subsec:eval-operation-type} \texttt{getattr} and \cref{subsec:eval-operation-class} \texttt{metadata}).
\end{itemize}

Further, all repositories include \texttt{README} files that describe how to install, configure, and test each system.

\subsection*{Requirements and dependencies}

\paragraph{System requirements}
\SYS and \chef were built and tested using \texttt{g++9.3.0} and \texttt{cmake-3.16}, and were successfully deployed in Ubuntu Server 20.04 LTS, CentOS 7.5, and CentOS 7.9 Linux distributions.
Operating system wise, the main requirements of these artifacts lie on the use of \texttt{LD\_PRELOAD} (\SYS) and \emph{UNIX Domain Sockets} (\SYS and \chef).

\paragraph{\SYS dependencies}
\SYS was built using the \PAIO v1.0.0 and \texttt{spdlog v1.8.1}\footnote{spdlog: \href{https://github.com/gabime/spdlog/tree/v1.8.1}{https://github.com/gabime/spdlog/tree/v1.8.1}} libraries. 
The former needs to be manually installed (installation steps are detailed in the \SYS repository), and the latter is installed at compile time (defined with a CMake rule).
Moreover, \SYS also uses \texttt{xoshiro-cpp}\footnote{Xoshiro-cpp: \href{https://github.com/Reputeless/Xoshiro-cpp}{https://github.com/Reputeless/Xoshiro-cpp}}, \texttt{tabulate}\footnote{tabulate: \href{https://github.com/p-ranav/tabulate}{https://github.com/p-ranav/tabulate}}, and \texttt{better-enums}\footnote{better-enums: \href{https://github.com/aantron/better-enums}{https://github.com/aantron/better-enums}} third-party libraries, which are embedded as single-header files.

\paragraph{\chef dependencies}
\chef was built using the \texttt{spdlog v1.8.1}, \texttt{grpc v1.37.0}\footnote{gRPC: \href{https://github.com/grpc/grpc/tree/v1.37.0}{https://github.com/grpc/grpc/tree/v1.37.0}}, \texttt{gflags v2.2.2}\footnote{gflags: \href{https://github.com/gflags/gflags/tree/v2.2.2}{https://github.com/gflags/gflags/tree/v2.2.2}}, \texttt{asio v1.18.0}\footnote{asio: \href{https://github.com/chriskohlhoff/asio/tree/asio-1-18-0}{https://github.com/chriskohlhoff/asio/tree/asio-1-18-0}}, and \texttt{yaml-cpp v0.6.3}\footnote{yaml-cpp: \href{https://github.com/jbeder/yaml-cpp/tree/yaml-cpp-0.6.3}{https://github.com/jbeder/yaml-cpp/tree/yaml-cpp-0.6.3}} libraries. 
All dependencies are installed at compile time, and all rules are defined at \chef's \texttt{CMakeLists.txt} file.

\subsection*{Commodity hardware experiments}

To ease the reproducibility of the paper experiments, we created a set of scripts to test the artifacts under a commodity hardware testbed.
For this scenario, we consider the \emph{per-operation type rate limiting} (\texttt{getattr}) and \emph{per-operation class rate limiting} (\texttt{metadata}) use cases, which are discussed in \cref{subsec:eval-operation-type} and \cref{subsec:eval-operation-class} of the paper, respectively.

\paragraph{Experimental testbed}
Experiments were conducted in a server equipped with a single 6-core Intel Core i5-9500 processor, 16 GiB of memory, and a Samsung NVMe SSD 970 EVO Plus 250 GiB disk.
Software-wise it used Ubuntu Server 20.04 LTS, with kernel 5.4.0 and an \texttt{ext4} file system.

\paragraph{Methodology}
All systems were executed under the same server. 
Experiments were conducted under the following steps:

\begin{enumerate}[nosep, leftmargin=*]
  \item Execute the \emph{global controller} using the script \texttt{lau\-nch\-\_co\-re\_con\-trol\-ler.sh}. 
  The \emph{global controller} will run at a user-level process, exposing a communication end-point (gRPC server) for \emph{local controllers} to connect.
  \item Execute the \emph{local controller} using the script \texttt{lau\-nch\_lo\-cal\_con\-trol\-ler.sh}.
  The \emph{local controller} will connect to the already running \emph{global controller}, while also creating a \emph{UNIX Domain Socket} for data plane stages to connect.
  \item Finally, execute the application (\emph{trace replayer}) with the script \texttt{lau\-nch\_pa\-dll\_ap\-pli\-ca\-ti\-on.sh}.
  The application will be spawned with a \texttt{LD\_PRELOAD} hook for the data plane stage (\emph{i.e.,} replace \texttt{libc.so} calls with those exposed by \SYS).
\end{enumerate} 

For the experiments described at \cref{subsec:eval-operation-type}, the \emph{trace replayer} replays the \texttt{get\-attr\_log.txt} file, while for \cref{subsec:eval-operation-class}'s experiments, it simultaneously replays \texttt{get\-at\-tr\_log.txt}, \texttt{re\-na\-me\_log.txt}, and \texttt{o\-pen\_log.txt} log files.\footnote{All logs can be found at \texttt{md\-re\-pla\-yer/lo\-gs/}.}
All traces are synthetic (\emph{i.e.,} not the original \abcifs logs discussed in \cref{sec:background}), but follow a variable metadata rate distribution.
Both experiments are executed over 1 minute, and \SYS adjusts the rate at which POSIX operations are submitted to the file system every 20 seconds.

\paragraph{Testing scripts} 
The scripts discussed in this section can be found at the \texttt{scri\-pts/com\-mo\-di\-ty\_scri\-pts} folder of the Zenodo public artifacts: 
\texttt{/per\_type\_getattr} for the \cref{subsec:eval-operation-type} and \texttt{/per\_class\_metadata} for the \cref{subsec:eval-operation-class} experiments.

\subsection*{Frontera experiments}
All experiments presented in the paper can be reproduced using the scripts available at \texttt{scripts/fron\-te\-ra\_scri\-pts} folder of the Zenodo public artifacts, which are ready to be executed at TACC's Frontera supercomputer.

\paragraph{Experimental testbed}
Experiments were conducted over two hardware configurations.
For the \cref{subsec:eval-operation-type}, \cref{subsec:eval-operation-class} \texttt{metadata}, and \cref{subsec:eval-job} experiments, we used compute nodes of Frontera's \texttt{normal} job queue, which are equipped with two 28-core Intel Xeon processors, 192 GiB of RAM, and a Mellanox InfiniBand HDR-100 network card, running CentOS 7.9. 
The production PFS is a Lustre file system composed of 4 MDSs, each with a single MDT, and 32 OST nodes with 22 PiB of storage capacity.
The \cref{subsec:eval-operation-class} \texttt{data} (TensorFlow) were executed at computes nodes of Frontera's \texttt{rtx} job queue, which are equipped with two 16-core Intel Xeon processors, 128 GiB of RAM, four NVIDIA Quadro RTX 5000 GPUs, and a Mellanox InfiniBand FDR network card, running CentOS 7.9. 

\paragraph{Workloads}
Metadata workloads were conducted using the \emph{trace replayer}.
All traces are synthetic (\emph{i.e.,} not the original \abcifs logs), but follow a variable metadata rate distribution.

Data experiments were twofold: (1) for \cref{subsec:eval-operation-type} \texttt{read} and \texttt{write} testing scenarios, we used IOR (commit \href{https://github.com/hpc/ior/tree/1076c8942caf22788e8b43045967bf349f10c34a}{\#1076c89}) sequential read/write workloads (IOR scripts can be found at \texttt{per\_ty\-pe/pa\-dll\_ior\_job.sh});
and (2) for the \cref{subsec:eval-operation-class} \texttt{data} experiment, we used TensorFlow v2.3.2 with the LeNet training model, configured with a batch size of 128 TFRecords, and ImageNet dataset  (TensorFlow scripts can be found at \texttt{per\_class/data/test\_tensorflow.sh}).\footnote{ImageNet: \href{https://www.image-net.org/challenges/LSVRC/2012/}{https://www.image-net.org/challenges/LSVRC/2012/}.}

\paragraph{Methodology}
For all experiments, the \emph{global controller} runs at a dedicated compute node (\texttt{core\_job.sh}).
For \cref{subsec:eval-operation-type} and \cref{subsec:eval-operation-class}, we use an additional compute node to host the \emph{local controller} (\texttt{local\_job.sh}), the application (\emph{trace replayer}), and the data plane stage. 

For \cref{subsec:eval-job} experiments, we use four additional compute nodes, each hosting a \emph{local controller}, the application (\emph{trace replayer}), and the data plane stage.

\paragraph{Testing scripts}
For each evaluation scenario, we used the following scripts:

\begin{itemize}
  \item \textbf{\cref{subsec:eval-operation-type}:} \texttt{fron\-te\-ra\_scri\-pts/per\_typ\-e}
  \begin{itemize}
    \item \textbf{\texttt{open}:} \texttt{launch\_open\_test.sh}
    \item \textbf{\texttt{getattr}:} \texttt{launch\_getattr\_test.sh}
    \item \textbf{\texttt{read}:} \texttt{launch\_read\_test.sh}
    \item \textbf{\texttt{write}:} \texttt{launch\_write\_test.sh}
  \end{itemize} 
  \item \textbf{\cref{subsec:eval-operation-class}:} \texttt{fron\-te\-ra\_scri\-pts/per\_cla\-ss}
  \begin{itemize}
    \item \textbf{\texttt{metadata}:} \texttt{metadata/launch\_test.sh}
    \item \textbf{\texttt{data}:} \texttt{data/launch\_test.sh}
  \end{itemize} 
  \item \textbf{\cref{subsec:eval-job}:} For the \cref{subsec:eval-job} experiments, we only provide scripts for the testing scenario \#1 (the remainder are just a variation of the load and job priorities).
  Further, we provide five subdirectories, each with the corresponding scripts for the different algorithms discussed in \cref{sec:control-algorihtms}, including: \emph{baseline}, \emph{uniform}, \emph{priority}, \emph{proportional sharing}, and \emph{PSFA}.
  All scripts can be found in the \texttt{fron\-te\-ra\_scri\-pts/per\_job} folder.
  \begin{itemize}
    \item \textbf{\texttt{baseline}:} \texttt{baseline/launch\_test.sh}
    \item \textbf{\texttt{uniform}:} \texttt{static/launch\_test.sh}
    \item \textbf{\texttt{priority}:} \texttt{priority/launch\_test.sh}
    \item \textbf{\texttt{psharing}:} \texttt{proportional/launch\_test.sh}
    \item \textbf{\texttt{psfa}:} \texttt{psfa/launch\_test.sh}
  \end{itemize} 
\end{itemize}

\end{document}